\documentclass{article}
\usepackage{epsfig}
\usepackage{slashbox}
\def\lapprox{\lower .7ex\hbox{$\;\stackrel{\textstyle <}{\sim}\;$}}
\def\gapprox{\lower .7ex\hbox{$\;\stackrel{\textstyle >}{\sim}\;$}}

\def\d{{\rm d}}

\def\sab{s_{12}}
\def\sac{s_{13}}
\def\sbc{s_{23}}
\def\sij{s_{ij}}
\def\sabc{s_{123}}
\def\sabcp{s^{\prime}_{123}}

\def\boxLO{
\mbox{\parbox{2.5cm}{\hspace{0.25cm}
\begin{picture}(2,1)
\thicklines
\put(0.2,0){\vector(-1,0){0.1}}
\put(1.8,0){\vector(1,0){0.1}}
\put(1.8,1){\vector(1,0){0.1}}
\put(0.3,1){\vector(1,0){0.1}}
\put(0,0){\line(1,0){2}}
\put(0,1){\line(1,0){2}}
\put(0.5,0){\line(0,1){1}}
\put(1.5,0){\line(0,1){1}}
\put(0.15,0.2){$p_1$}
\put(1.7,0.2){$p_3$}
\put(1.7,0.7){$p_2$}
\put(0.15,0.7){$q$}
\end{picture}
}} 
\hfill}

\def\boxLOpdmc{
\mbox{\parbox{2.5cm}{\hspace{0.25cm}
\begin{picture}(2,1)
\thicklines
\put(0.2,0){\vector(-1,0){0.1}}
\put(1.8,0){\vector(1,0){0.1}}
\put(1.8,1){\vector(1,0){0.1}}
\put(0.3,1){\vector(1,0){0.1}}
\put(1,1){\circle*{0.15}}
\put(1.368,0.412){$\times$}
\put(0,0){\line(1,0){2}}
\put(0,1){\line(1,0){2}}
\put(0.5,0){\line(0,1){1}}
\put(1.5,0){\line(0,1){1}}
\put(0.15,0.2){$p_1$}
\put(1.7,0.2){$p_3$}
\put(1.7,0.7){$p_2$}
\put(0.15,0.7){$q$}
\end{picture}
}} 
\hfill}

\def\boxLOpamb{
\mbox{\parbox{2.5cm}{\hspace{0.25cm}
\begin{picture}(2,1)
\thicklines
\put(0.2,0){\vector(-1,0){0.1}}
\put(1.8,0){\vector(1,0){0.1}}
\put(1.8,1){\vector(1,0){0.1}}
\put(0.3,1){\vector(1,0){0.1}}
\put(0.5,0.5){\circle*{0.15}}
\put(0.87,-0.085){$\times$}
\put(0,0){\line(1,0){2}}
\put(0,1){\line(1,0){2}}
\put(0.5,0){\line(0,1){1}}
\put(1.5,0){\line(0,1){1}}
\put(0.15,0.2){$p_1$}
\put(1.7,0.2){$p_3$}
\put(1.7,0.7){$p_2$}
\put(0.15,0.7){$q$}
\end{picture}
}} 
\hfill}

\newcommand{\bubbleLO}[1]{
\mbox{\parbox{2.5cm}{\hspace{0.25cm}
\begin{picture}(2,1)
\thicklines
\put(0.3,0.5){\vector(1,0){0.1}}
\put(0,0.5){\line(1,0){0.5}}
\put(1.5,0.5){\line(1,0){0.5}}
\put(1,0.5){\circle{1}}
\put(0.25,0.7){\makebox(0,0)[b]{$#1$}}
\end{picture}
}}
\hfill}

\newcommand{\bubbleNLO}[1]{
\mbox{\parbox{2.5cm}{\hspace{0.25cm}
\begin{picture}(2,1)
\thicklines
\put(0.3,0.5){\vector(1,0){0.1}}
\put(0,0.5){\line(1,0){2}}
\put(1,0.5){\circle{1}}
\put(0.25,0.7){\makebox(0,0)[b]{$#1$}}
\end{picture}
}}
\hfill}

\newcommand{\bubblexNLO}[1]{
\mbox{\parbox{2.5cm}{\hspace{0.25cm}
\begin{picture}(2,1)
\thicklines
\put(0.3,0.5){\vector(1,0){0.1}}
\put(0,0.5){\line(1,0){0.5}}
\put(1.5,0.5){\line(1,0){0.5}}
\put(1.5,1){\oval(1,1)[bl]}
\put(1,0.5){\circle{1}}
\put(0.25,0.7){\makebox(0,0)[b]{$#1$}}
\end{picture}
}}
\hfill}

\newcommand{\doublebubbleNLO}[1]{
\mbox{\parbox{3.5cm}{\hspace{0.25cm}
\begin{picture}(3,1)
\thicklines
\put(0.3,0.5){\vector(1,0){0.1}}
\put(0,0.5){\line(1,0){0.5}}
\put(2.5,0.5){\line(1,0){0.5}}
\put(1,0.5){\circle{1}}
\put(2,0.5){\circle{1}}
\put(0.25,0.7){\makebox(0,0)[b]{$#1$}}
\end{picture}
}}
\hfill}

\newcommand{\doublebubblexNLO}[3]{
\mbox{\parbox{3.5cm}{\hspace{0.25cm}
\begin{picture}(3,1.5)
\thicklines
\put(0.3,0.5){\vector(1,0){0.1}}
\put(2.75,0.5){\vector(1,0){0.1}}
\put(1.5,1.0){\vector(0,1){0.1}}
\put(0,0.5){\line(1,0){0.5}}
\put(1.5,0.5){\line(0,1){0.6}}
\put(2.5,0.5){\line(1,0){0.5}}
\put(1,0.5){\circle{1}}
\put(2,0.5){\circle{1}}
\put(0.25,0.7){\makebox(0,0)[b]{$#1$}}
\put(2.75,0.7){\makebox(0,0)[b]{$#2$}}
\put(1.5,1.25){\makebox(0,0)[b]{$#3$}}
\end{picture}
}}
\hfill}

\newcommand{\bubblecrossNLO}[1]{
\mbox{\parbox{2.5cm}{\hspace{0.25cm}
\begin{picture}(2,1)
\thicklines
\put(0.3,0.5){\vector(1,0){0.1}}
\put(0,0.5){\line(1,0){0.5}}
\put(1.5,0.5){\line(1,0){0.5}}
\put(1,1){\line(0,-1){1}}
\put(1,0.5){\circle{1}}
\put(0.25,0.7){\makebox(0,0)[b]{$#1$}}
\end{picture}
}}
\hfill}

\newcommand{\trianglexNLO}[3]{
\mbox{\parbox{3cm}{\hspace{0.25cm}
\begin{picture}(2.5,1.4)
\thicklines
\put(0.3,0.7){\vector(1,0){0.1}}
\put(1.7,0.2){\vector(1,0){0.1}}
\put(1.7,1.2){\vector(1,0){0.1}}
\put(0,0.7){\line(1,0){0.5}}
\put(0.5,1.2){\oval(1,1)[br]}
\put(1,1.2){\line(1,0){1}}
\put(1,0.2){\line(1,0){1}}
\put(1,0.7){\circle{1}}
\put(0.25,0.9){\makebox(0,0)[b]{$#1$}}
\put(2.05,1.2){\makebox(0,0)[l]{$#2$}}
\put(2.05,0.2){\makebox(0,0)[l]{$#3$}}
\end{picture}
}}
\hfill}

\newcommand{\triangleNLO}[3]{
\mbox{\parbox{3cm}{\hspace{0.25cm}
\begin{picture}(2.5,1.4)
\thicklines
\put(0.3,0.7){\vector(1,0){0.1}}
\put(1.7,0.2){\vector(1,0){0.1}}
\put(1.7,1.2){\vector(1,0){0.1}}
\put(0,0.7){\line(1,0){0.5}}
\put(1,1.2){\line(0,-1){1}}
\put(1,1.2){\line(1,0){1}}
\put(1,0.2){\line(1,0){1}}
\put(1,0.7){\circle{1}}
\put(0.25,0.9){\makebox(0,0)[b]{$#1$}}
\put(2.05,1.2){\makebox(0,0)[l]{$#2$}}
\put(2.05,0.2){\makebox(0,0)[l]{$#3$}}
\end{picture}
}}
\hfill}

\newcommand{\triangleaNLO}[3]{
\mbox{\parbox{3cm}{\hspace{0.25cm}
\begin{picture}(2.5,1.4)
\thicklines
\put(0.3,0.7){\vector(1,0){0.1}}
\put(1.9,0.2){\vector(1,0){0.1}}
\put(1.9,1.2){\vector(1,0){0.1}}
\put(0,0.7){\line(1,0){0.5}}
\put(0.5,0.7){\line(2,1){1}}
\put(0.5,0.7){\line(2,-1){1}}
\put(1.5,1.2){\line(0,-1){1}}
\put(1.5,1.2){\line(1,0){0.5}}
\put(1.5,0.2){\line(1,0){0.5}}
\put(0.5,0.7){\line(1,0){1}}
\put(0.25,0.9){\makebox(0,0)[b]{$#1$}}
\put(2.05,1.2){\makebox(0,0)[l]{$#2$}}
\put(2.05,0.2){\makebox(0,0)[l]{$#3$}}
\end{picture}
}}
\hfill}

\newcommand{\trianglebNLO}[3]{
\mbox{\parbox{3cm}{\hspace{0.25cm}
\begin{picture}(2.5,1.4)
\thicklines
\put(0.3,0.7){\vector(1,0){0.1}}
\put(1.9,0.2){\vector(1,0){0.1}}
\put(1.9,1.2){\vector(1,0){0.1}}
\put(0,0.7){\line(1,0){0.5}}
\put(0.5,0.7){\line(2,1){1}}
\put(0.5,0.7){\line(2,-1){1}}
\put(1.5,1.2){\line(0,-1){1}}
\put(1.5,1.2){\line(1,0){0.5}}
\put(1.5,0.2){\line(1,0){0.5}}
\put(1.5,0.2){\line(-1,2){0.4}}
\put(0.25,0.9){\makebox(0,0)[b]{$#1$}}
\put(2.05,1.2){\makebox(0,0)[l]{$#2$}}
\put(2.05,0.2){\makebox(0,0)[l]{$#3$}}
\end{picture}
}}
\hfill}

\newcommand{\trianglecNLO}[3]{
\mbox{\parbox{3cm}{\hspace{0.25cm}
\begin{picture}(2.5,1.4)
\thicklines
\put(0.3,0.7){\vector(1,0){0.1}}
\put(1.9,0.2){\vector(1,0){0.1}}
\put(1.9,1.2){\vector(1,0){0.1}}
\put(0,0.7){\line(1,0){0.5}}
\put(0.5,0.7){\line(2,-1){1}}
\put(0.5,0.7){\line(1,1){0.5}}
\put(1.5,1.2){\line(0,-1){1}}
\put(1.5,1.2){\line(1,0){0.5}}
\put(1.5,0.2){\line(1,0){0.5}}
\put(1.25,1.2){\circle{0.5}}
\put(0.25,0.9){\makebox(0,0)[b]{$#1$}}
\put(2.05,1.2){\makebox(0,0)[l]{$#2$}}
\put(2.05,0.2){\makebox(0,0)[l]{$#3$}}
\end{picture}
}}
\hfill}

\newcommand{\triangledNLO}[3]{
\mbox{\parbox{3cm}{\hspace{0.25cm}
\begin{picture}(2.5,1.4)
\thicklines
\put(0.3,0.2){\vector(1,0){0.1}}
\put(1.9,0.2){\vector(1,0){0.1}}
\put(1.9,1.2){\vector(1,0){0.1}}
\put(0,0.2){\line(1,0){0.5}}
\put(0.5,0.2){\line(1,1){1}}
\put(0.5,0.2){\line(1,0){1}}
\put(1.5,1.2){\line(0,-1){0.5}}
\put(1.5,1.2){\line(1,0){0.5}}
\put(1.5,0.2){\line(1,0){0.5}}
\put(1.5,0.45){\circle{0.5}}
\put(0.25,0.4){\makebox(0,0)[b]{$#1$}}
\put(2.05,1.2){\makebox(0,0)[l]{$#2$}}
\put(2.05,0.2){\makebox(0,0)[l]{$#3$}}
\end{picture}
}}
\hfill}

\newcommand{\boxbubbleapNLO}[4]{
\mbox{\parbox{4cm}{\hspace{0.25cm}
\begin{picture}(3.5,1.4)
\thicklines
\put(0.7,0.2){\vector(-1,0){0.1}}
\put(2.8,0.2){\vector(1,0){0.1}}
\put(2.8,1.2){\vector(1,0){0.1}}
\put(0.8,1.2){\vector(1,0){0.1}}
\put(0.5,0.2){\line(1,0){2.5}}
\put(0.5,1.2){\line(1,0){2.5}}
\put(1,0.2){\line(0,1){1}}
\put(2,1.2){\line(0,-1){0.5}}
\put(2,0.45){\circle{0.5}}
\put(0.45,1.2){\makebox(0,0)[r]{$#1$}}
\put(0.45,0.2){\makebox(0,0)[r]{$#2$}}
\put(3.05,1.2){\makebox(0,0)[l]{$#3$}}
\put(3.05,0.2){\makebox(0,0)[l]{$#4$}}
\end{picture}
}} 
\hfill}

\newcommand{\boxbubbleaNLO}[4]{
\mbox{\parbox{4cm}{\hspace{0.25cm}
\begin{picture}(3.5,1.4)
\thicklines
\put(0.7,0.2){\vector(-1,0){0.1}}
\put(2.8,0.2){\vector(1,0){0.1}}
\put(2.8,1.2){\vector(1,0){0.1}}
\put(0.8,1.2){\vector(1,0){0.1}}
\put(0.5,0.2){\line(1,0){2.5}}
\put(0.5,1.2){\line(1,0){2.5}}
\put(1,0.2){\line(0,1){1}}
\put(2,0.7){\circle{1}}
\put(0.45,1.2){\makebox(0,0)[r]{$#1$}}
\put(0.45,0.2){\makebox(0,0)[r]{$#2$}}
\put(3.05,1.2){\makebox(0,0)[l]{$#3$}}
\put(3.05,0.2){\makebox(0,0)[l]{$#4$}}
\end{picture}
}} 
\hfill}

\newcommand{\boxbubblebNLO}[4]{
\mbox{\parbox{4cm}{\hspace{0.25cm}
\begin{picture}(3.5,1.4)
\thicklines
\put(0.7,0.2){\vector(-1,0){0.1}}
\put(2.8,0.2){\vector(1,0){0.1}}
\put(2.8,1.2){\vector(1,0){0.1}}
\put(0.8,1.2){\vector(1,0){0.1}}
\put(0.5,0.2){\line(1,0){2.5}}
\put(0.5,1.2){\line(1,0){2.5}}
\put(2.5,0.2){\line(0,1){1}}
\put(1.5,0.7){\circle{1}}
\put(0.45,1.2){\makebox(0,0)[r]{$#1$}}
\put(0.45,0.2){\makebox(0,0)[r]{$#2$}}
\put(3.05,1.2){\makebox(0,0)[l]{$#3$}}
\put(3.05,0.2){\makebox(0,0)[l]{$#4$}}
\end{picture}
}} 
\hfill}

\newcommand{\boxbubblebpNLO}[4]{
\mbox{\parbox{4cm}{\hspace{0.25cm}
\begin{picture}(3.5,1.4)
\thicklines
\put(0.7,0.2){\vector(-1,0){0.1}}
\put(2.8,0.2){\vector(1,0){0.1}}
\put(2.8,1.2){\vector(1,0){0.1}}
\put(0.8,1.2){\vector(1,0){0.1}}
\put(0.5,0.2){\line(1,0){2.5}}
\put(0.5,1.2){\line(1,0){2.5}}
\put(2.5,0.2){\line(0,1){1}}
\put(1.5,1.2){\line(0,-1){0.5}}
\put(1.5,0.45){\circle{0.5}}
\put(0.45,1.2){\makebox(0,0)[r]{$#1$}}
\put(0.45,0.2){\makebox(0,0)[r]{$#2$}}
\put(3.05,1.2){\makebox(0,0)[l]{$#3$}}
\put(3.05,0.2){\makebox(0,0)[l]{$#4$}}
\end{picture}
}} 
\hfill}

\newcommand{\boxxaNLO}[4]{
\mbox{\parbox{3.5cm}{\hspace{0.25cm}
\begin{picture}(3,1.4)
\thicklines
\put(0.7,0.2){\vector(-1,0){0.1}}
\put(2.3,0.2){\vector(1,0){0.1}}
\put(2.3,1.2){\vector(1,0){0.1}}
\put(0.8,1.2){\vector(1,0){0.1}}
\put(0.5,0.2){\line(1,0){2}}
\put(2,0.2){\line(-1,1){1}}
\put(0.5,1.2){\line(1,0){2}}
\put(1,0.2){\line(0,1){1}}
\put(2,0.2){\line(0,1){1}}
\put(0.45,1.2){\makebox(0,0)[r]{$#1$}}
\put(0.45,0.2){\makebox(0,0)[r]{$#2$}}
\put(2.55,1.2){\makebox(0,0)[l]{$#3$}}
\put(2.55,0.2){\makebox(0,0)[l]{$#4$}}
\end{picture}
}} 
\hfill}

\newcommand{\boxxapNLO}[4]{
\mbox{\parbox{3.5cm}{\hspace{0.25cm}
\begin{picture}(3,1.4)
\thicklines
\put(0.7,0.2){\vector(-1,0){0.1}}
\put(2.3,0.2){\vector(1,0){0.1}}
\put(2.3,1.2){\vector(1,0){0.1}}
\put(0.8,1.2){\vector(1,0){0.1}}
\put(0.5,0.2){\line(1,0){2}}
\put(2,0.2){\line(-1,2){0.5}}
\put(0.5,1.2){\line(1,0){2}}
\put(1,0.2){\line(0,1){1}}
\put(2,0.2){\line(0,1){1}}
\put(0.45,1.2){\makebox(0,0)[r]{$#1$}}
\put(0.45,0.2){\makebox(0,0)[r]{$#2$}}
\put(2.55,1.2){\makebox(0,0)[l]{$#3$}}
\put(2.55,0.2){\makebox(0,0)[l]{$#4$}}
\end{picture}
}} 
\hfill}

\newcommand{\boxxamNLO}[4]{
\mbox{\parbox{3.5cm}{\hspace{0.25cm}
\begin{picture}(3,1.4)
\thicklines
\put(0.7,0.2){\vector(-1,0){0.1}}
\put(2.3,0.2){\vector(1,0){0.1}}
\put(2.3,1.2){\vector(1,0){0.1}}
\put(0.8,1.2){\vector(1,0){0.1}}
\put(0.5,0.2){\line(1,0){2}}
\put(1.5,0.2){\line(-1,2){0.5}}
\put(0.5,1.2){\line(1,0){2}}
\put(1,0.2){\line(0,1){1}}
\put(2,0.2){\line(0,1){1}}
\put(0.45,1.2){\makebox(0,0)[r]{$#1$}}
\put(0.45,0.2){\makebox(0,0)[r]{$#2$}}
\put(2.55,1.2){\makebox(0,0)[l]{$#3$}}
\put(2.55,0.2){\makebox(0,0)[l]{$#4$}}
\end{picture}
}} 
\hfill}

\newcommand{\boxxbNLO}[4]{
\mbox{\parbox{3.5cm}{\hspace{0.25cm}
\begin{picture}(3,1.4)
\thicklines
\put(0.7,0.2){\vector(-1,0){0.1}}
\put(2.3,0.2){\vector(1,0){0.1}}
\put(2.3,1.2){\vector(1,0){0.1}}
\put(0.8,1.2){\vector(1,0){0.1}}
\put(0.5,0.2){\line(1,0){2}}
\put(1,0.2){\line(1,1){1}}
\put(0.5,1.2){\line(1,0){2}}
\put(1,0.2){\line(0,1){1}}
\put(2,0.2){\line(0,1){1}}
\put(0.45,1.2){\makebox(0,0)[r]{$#1$}}
\put(0.45,0.2){\makebox(0,0)[r]{$#2$}}
\put(2.55,1.2){\makebox(0,0)[l]{$#3$}}
\put(2.55,0.2){\makebox(0,0)[l]{$#4$}}
\end{picture}
}} 
\hfill}

\newcommand{\boxxbmNLO}[4]{
\mbox{\parbox{3.5cm}{\hspace{0.25cm}
\begin{picture}(3,1.4)
\thicklines
\put(0.7,0.2){\vector(-1,0){0.1}}
\put(2.3,0.2){\vector(1,0){0.1}}
\put(2.3,1.2){\vector(1,0){0.1}}
\put(0.8,1.2){\vector(1,0){0.1}}
\put(0.5,0.2){\line(1,0){2}}
\put(1.5,0.2){\line(1,2){0.5}}
\put(0.5,1.2){\line(1,0){2}}
\put(1,0.2){\line(0,1){1}}
\put(2,0.2){\line(0,1){1}}
\put(0.45,1.2){\makebox(0,0)[r]{$#1$}}
\put(0.45,0.2){\makebox(0,0)[r]{$#2$}}
\put(2.55,1.2){\makebox(0,0)[l]{$#3$}}
\put(2.55,0.2){\makebox(0,0)[l]{$#4$}}
\end{picture}
}} 
\hfill}

\newcommand{\boxxbmcrossNLO}[4]{
\mbox{\parbox{4.5cm}{\hspace{0.25cm}
\begin{picture}(4,1.4)
\thicklines
\put(0.7,0.2){\vector(-1,0){0.1}}
\put(2.3,0.2){\vector(1,0){0.1}}
\put(2.3,1.2){\vector(1,0){0.1}}
\put(0.8,1.2){\vector(1,0){0.1}}
\put(0.5,0.2){\line(1,0){2}}
\put(1.5,0.2){\line(1,2){0.2}}
\put(2,1.2){\line(-1,-2){0.2}}
\put(0.5,1.2){\line(1,0){2}}
\put(1,0.2){\line(1,2){0.5}}
\put(1.5,1.2){\line(1,-2){0.5}}
\put(0.45,1.2){\makebox(0,0)[r]{$#1$}}
\put(0.45,0.2){\makebox(0,0)[r]{$#2$}}
\put(2.55,1.2){\makebox(0,0)[l]{$#3$}}
\put(2.55,0.2){\makebox(0,0)[l]{$#4$}}
\end{picture}
}} 
\hfill}

\newcommand{\boxxbdotNLO}[4]{
\mbox{\parbox{3.5cm}{\hspace{0.25cm}
\begin{picture}(3,1.4)
\thicklines
\put(0.7,0.2){\vector(-1,0){0.1}}
\put(2.3,0.2){\vector(1,0){0.1}}
\put(2.3,1.2){\vector(1,0){0.1}}
\put(0.8,1.2){\vector(1,0){0.1}}
\put(0.5,0.2){\line(1,0){2}}
\put(1,0.2){\line(1,1){1}}
\put(1.5,0.7){\circle*{0.2}}
\put(0.5,1.2){\line(1,0){2}}
\put(1,0.2){\line(0,1){1}}
\put(2,0.2){\line(0,1){1}}
\put(0.45,1.2){\makebox(0,0)[r]{$#1$}}
\put(0.45,0.2){\makebox(0,0)[r]{$#2$}}
\put(2.55,1.2){\makebox(0,0)[l]{$#3$}}
\put(2.55,0.2){\makebox(0,0)[l]{$#4$}}
\end{picture}
}} 
\hfill}

\newcommand{\triaplanNLO}[3]{
\mbox{\parbox{3cm}{\hspace{0.25cm}
\begin{picture}(2.5,1.4)
\thicklines
\put(0.3,0.7){\vector(1,0){0.1}}
\put(1.9,0.2){\vector(1,0){0.1}}
\put(1.9,1.2){\vector(1,0){0.1}}
\put(0,0.7){\line(1,0){0.5}}
\put(0.5,0.7){\line(1,1){0.5}}
\put(0.5,0.7){\line(1,-1){0.5}}
\put(1.5,1.2){\line(0,-1){1}}
\put(1.0,1.2){\line(0,-1){1}}
\put(1,1.2){\line(1,0){1}}
\put(1,0.2){\line(1,0){1}}
\put(0.25,0.9){\makebox(0,0)[b]{$#1$}}
\put(2.05,1.2){\makebox(0,0)[l]{$#2$}}
\put(2.05,0.2){\makebox(0,0)[l]{$#3$}}
\end{picture}
}}
\hfill}

\newcommand{\triaplanxNLO}[3]{
\mbox{\parbox{3cm}{\hspace{0.25cm}
\begin{picture}(2.5,1.4)
\thicklines
\put(0.3,0.7){\vector(1,0){0.1}}
\put(1.9,0.2){\vector(1,0){0.1}}
\put(1.9,1.2){\vector(1,0){0.1}}
\put(0,0.7){\line(1,0){0.5}}
\put(0.5,0.7){\line(2,1){1}}
\put(0.5,0.7){\line(2,-1){1}}
\put(1.5,1.2){\line(0,-1){1}}
\put(1.5,1.2){\line(1,0){0.5}}
\put(1.5,0.2){\line(1,0){0.5}}
\put(1.5,0.7){\line(-2,-1){0.5}}
\put(0.25,0.9){\makebox(0,0)[b]{$#1$}}
\put(2.05,1.2){\makebox(0,0)[l]{$#2$}}
\put(2.05,0.2){\makebox(0,0)[l]{$#3$}}
\end{picture}
}}
\hfill}

\newcommand{\boxtriaaNLO}[4]{
\mbox{\parbox{3.5cm}{\hspace{0.25cm}
\begin{picture}(3,1.4)
\thicklines
\put(0.7,0.2){\vector(-1,0){0.1}}
\put(2.3,0.2){\vector(1,0){0.1}}
\put(2.3,1.2){\vector(1,0){0.1}}
\put(0.8,1.2){\vector(1,0){0.1}}
\put(0.5,0.2){\line(1,0){2}}
\put(1.5,1.2){\line(1,-1){0.5}}
\put(0.5,1.2){\line(1,0){2}}
\put(1,0.2){\line(0,1){1}}
\put(2,0.2){\line(0,1){1}}
\put(0.45,1.2){\makebox(0,0)[r]{$#1$}}
\put(0.45,0.2){\makebox(0,0)[r]{$#2$}}
\put(2.55,1.2){\makebox(0,0)[l]{$#3$}}
\put(2.55,0.2){\makebox(0,0)[l]{$#4$}}
\end{picture}
}} 
\hfill}

\newcommand{\boxtriabNLO}[4]{
\mbox{\parbox{3.5cm}{\hspace{0.25cm}
\begin{picture}(3,1.4)
\thicklines
\put(0.7,0.2){\vector(-1,0){0.1}}
\put(2.3,0.2){\vector(1,0){0.1}}
\put(2.3,1.2){\vector(1,0){0.1}}
\put(0.8,1.2){\vector(1,0){0.1}}
\put(0.5,0.2){\line(1,0){2}}
\put(1.5,0.2){\line(1,1){0.5}}
\put(0.5,1.2){\line(1,0){2}}
\put(1,0.2){\line(0,1){1}}
\put(2,0.2){\line(0,1){1}}
\put(0.45,1.2){\makebox(0,0)[r]{$#1$}}
\put(0.45,0.2){\makebox(0,0)[r]{$#2$}}
\put(2.55,1.2){\makebox(0,0)[l]{$#3$}}
\put(2.55,0.2){\makebox(0,0)[l]{$#4$}}
\end{picture}
}} 
\hfill}

\begin{document}
\unitlength1cm
\begin{titlepage}
\renewcommand{\thefootnote}{\fnsymbol{footnote}}
\vspace*{-1cm}
\begin{flushright}
TTP99--49\\
December 1999 
\end{flushright}                                
\vskip 3.5cm
\begin{center}
{\Large\bf Differential Equations for Two-Loop Four-Point Functions}
\vskip 1.cm
{\large  T.~Gehrmann} and {\large E.~Remiddi}\footnote{Supported by
    Alexander-von-Humboldt
    Stiftung, permanent address: Dipartimento di Fisica,
    Universit\`{a} di Bologna, I-40126 Bologna, Italy} 
\vskip .7cm
{\it Institut f\"ur Theoretische Teilchenphysik,
Universit\"at Karlsruhe, D-76128 Karlsruhe, Germany}
\end{center}
\vskip 2.6cm

\begin{abstract}
At variance with fully inclusive quantities, which have been computed 
already at the two- or three-loop level, most exclusive observables are 
still known only at one-loop, as further progress was hampered 
so far by the greater computational problems encountered in the study 
of multi-leg amplitudes beyond one loop. 
We show in this paper how the use of tools already employed in inclusive 
calculations can be suitably extended to the 
computation of loop integrals appearing in the virtual corrections to 
exclusive observables, namely two-loop four-point
functions with massless propagators and up to one off-shell leg. 
We find that multi-leg integrals, in addition to integration-by-parts 
identities, obey also identities resulting from Lorentz-invariance. The
combined set of these identities can be used to reduce the large number of
integrals appearing in an actual calculation to a small number of
master integrals. 
We then write down explicitly the differential equations in the external 
invariants fulfilled by these master integrals, 
and point out that the equations can be used 
as an efficient method of evaluating the master integrals themselves. 
We outline strategies for the solution of the differential equations, 
and demonstrate the application of the method on several examples. 
\end{abstract}
\vfill
\end{titlepage}                                                                
\newpage                                                                       

\renewcommand{\theequation}{\mbox{\arabic{section}.\arabic{equation}}}
\section{Introduction}
\setcounter{equation}{0}
\setcounter{footnote}{0}
\renewcommand{\thefootnote}{\arabic{footnote}}
Perturbative corrections to many inclusive quantities have been computed 
to the two- and three-loop level in past years. From the technical 
point of view, these inclusive calculations correspond to the computation of 
multi-loop two-point functions, for which many elaborate
calculational tools have been developed. In contrast, 
corrections to exclusive
observables, such as jet production rates, could up to now only be
computed at the one-loop level. These calculations require the 
computation of multi-leg amplitudes to the required number of loops,
which beyond the one-loop level 
turn out to be a calculational challenge obstructing further progress. 
Despite considerable progress made in recent times, many of the
two-loop integrals relevant for the calculation of jet observables
beyond next-to-leading order are still unknown. One particular class of
yet unknown
integrals appearing in the two-loop corrections to three jet production
in electron-positron collisions, to two-plus-one jet production in 
electron-proton collisions and to vector boson plus jet production in 
proton-proton collisions are two-loop four-point functions with massless 
internal propagators and one external leg off-shell. 

Taking two-loop four-point 
integrals arising in the calculation of Feynman diagrams in covariant 
gauges (non-covariant gauges introduce integrals of a structure beyond the 
treatment of this paper)
as an example, we elaborate on in this paper several
techniques to compute multi-leg amplitudes beyond one loop. We
demonstrate how integration-by-parts identities (already known to be a very
valuable tool in inclusive calculations) and identities following from
Lorentz-invariance (which are non-trivial only for integrals depending 
on at least 
two independent external momenta) can be used to reduce the large number 
of different integrals appearing in an actual calculation to a small 
number of master integrals. This reduction can be carried out
mechanically (by means of a small chain of 
computer programs), without
explicit reference to the actual structure of the integrals under
consideration and can also be used for the reduction of 
tensor integrals beyond one loop. 

The master integrals themselves, however, can not be computed from
these identities. We derive differential equations in the external
momenta for them. Solving these
differential equations, it is possible to compute the master integrals
without explicitly carrying out any loop integration, so that this
technique appears to be a valuable alternative to conventional
approaches for the computation of multi-loop integrals. 

The plan of the paper is as follows. In Section~\ref{sec:master}
we review the derivation of the integration by parts (IBP) identities 
 and introduce the Lorentz invariance (LI) identities. 
In Section~\ref{sec:diffeq} the differential equations for the master 
amplitudes are obtained. 
The practical application of these tools is outlined in detail 
in Section~\ref{sec:oneloop} on a self-contained rederivation of the 
one-loop massless box integral with one off-shell
leg.  
Section~\ref{sec:twoloop} contains examples of  
massless two-loop four-point functions with one off-shell
leg, evaluated for arbitrary space-time dimension.
 We show which of these functions can be reduced to simpler
functions and which are genuine master integrals, and compute some of
the master integrals by solving the corresponding differential
equations. Finally, Section~\ref{sec:conc} contains 
conclusions and an outlook on
potential future applications of the tools developed here. The 
higher transcendental functions appearing in our results for the 
one-loop and two-loop
integrals are summarised in an Appendix, where we also discuss how these 
functions can be expanded around the physical number of space-time
dimensions. 

\section{Reduction to Master Integrals}
\label{sec:master}
\setcounter{equation}{0}
Any scalar massless two-loop integral can be brought into the form 
\begin{equation}
I(p_1,\ldots,p_n) = \int \frac{\d^d k}{(2\pi)^d}\frac{\d^d l}{(2\pi)^d} 
\frac{1}{D_1^{m_1}\ldots D_{t}^{m_t}} S_1^{n_1} 
\ldots S_q^{n_q} \; ,
\label{eq:generic}
\end{equation}
where the $D_i$ are massless scalar propagators, depending on $k$, $l$ and the 
external momenta $p_1,\ldots,p_n$ while $S_i$ are scalar products
of a loop momentum with an external momentum or of the two loop
momenta.  The topology (interconnection of
propagators and external momenta) of the integral is uniquely 
determined by specifying the set $(D_1,\ldots,D_t)$
of $t$ different propagators in the graph. The integral itself is then
specified by the powers $m_i$ of all propagators and by the selection 
$(S_1,\ldots,S_q)$  of scalar products and their powers $(n_1,\ldots,n_q)$ . 
(all the $m_i$ are positive integers greater or equal to 1, while the 
$n_i$ are greater or equal to 0). 
Integrals of the same topology with the same dimension $r=\sum_i m_i$ 
of the denominator and same total number $s=\sum_i n_i$ of scalar products 
are denoted as a class of integrals $I_{t,r,s}$. The integration measure and 
scalar products appearing the above expression are in Minkowskian space, 
with the usual causal prescription for all propagators. The loop
integrations are carried out for arbitrary space-time dimension $d$, 
which acts as a regulator for divergencies appearing due to the
ultraviolet or infrared behaviour of the integrand (dimensional 
regularisation, \cite{dreg,hv}).

Any four-point function depends on three linearly independent 
external momenta, $p_1$, $p_2$ and $p_3$. At the two-loop level, one can 
combine the two loop momenta $k$ and $l$ and these external momenta 
to form 9 different scalar products involving $k$
or $l$. As the propagators present in the graph are (linearly independent) 
combinations of scalar products, only $9-t$
different scalar products can appear explicitly in an integral with $t$
different propagators. Since a two-loop four-point function can have at 
most seven different propagators, as can be found by considering the 
insertion of a propagator into a one-loop four-point function, one has 
in general $t\leq 7$, while the 
minimum number of massless propagators in a two-loop graph is $t=3$, 
corresponding to a two-point function. 
The number of different two-loop four-point 
integrals for given $t$ (number of different propagators), $r$ 
(sum of powers of all propagators) and $s$ (sum of powers of all 
scalar products) can be computed from simple combinatorics:
\begin{equation}
N(I_{t,r,s}) = { r-1 \choose r-t} {8-t+s \choose s} \; .
\label{eq:intnum}
\end{equation} 

The number $N(I_{t,r,s})$ of the integrals grows quickly as $r, s$ 
increase, but the integrals are related among themselves 
by various identities. 
One class of identities follows from the fact that the integral over the 
total derivative with respect to any loop momentum vanishes in
dimensional regularisation
\begin{equation}
\int \frac{\d^d k}{(2\pi)^d} \frac{\partial}{\partial k^{\mu}}
J(k,\ldots)  = 0,
\end{equation} 
where $J$ is any combination of propagators, scalar products
and loop momentum vectors. $J$ can be a vector or tensor of any rank. 
The resulting identities~\cite{hv,chet} are called integration-by-parts (IBP)
identities and can for two-loop integrals be cast into the form 
\begin{eqnarray}
\int \frac{\d^d k}{(2\pi)^d} \frac{\d^d l}{(2\pi)^d}
\frac{\partial}{\partial k^{\mu}} v^{\mu} f(k,l,p_i) & = & 0, \nonumber \\
\int \frac{\d^d k}{(2\pi)^d} \frac{\d^d l}{(2\pi)^d}
\frac{\partial}{\partial l^{\mu}} v^{\mu} f(k,l,p_i) & = & 0, 
\end{eqnarray}
where the integrand $f(k,l,p_i)$ is a scalar function, containing 
propagators and
scalar products and $v_{\mu}$ can be any external or loop momentum vector.
As a consequence, one obtains for a graph with $m$ loops and $n$
independent external momenta a total number of $N_{{\rm IBP}} = m(n+m)$. 
For a two-loop four-point function, this results in ten IBP identities for
each integrand. 

In addition to the IBP identities, one can also exploit the fact that
all integrals under consideration are Lorentz scalars (or, perhaps 
more precisely, ``$d$-rotational'' scalars) , which are
invariant under a Lorentz (or $d$-rotational) transformation of the 
external momenta. In
order to derive the resulting Lorentz invariance (LI) identities, we
consider an infinitesimal Lorentz transformation
\begin{equation}
p^{\mu} \to p^{\mu} + \delta p^{\mu} = 
p^{\mu} + \delta \epsilon^{\mu}_{\nu} p^{\nu} \qquad 
\mbox{with} \qquad \delta \epsilon^{\mu}_{\nu} = - \delta
\epsilon^{\nu}_{\mu}\;,
\end{equation}
which should not change the scalar Feynman integral
\begin{equation}
 I(p_1+\delta p_1,
\ldots , p_n+\delta p_n)= 
I(p_1,\ldots,p_n) \; .
\end{equation}
Expanding
\begin{equation}
 I(p_1+\delta p_1,
\ldots , p_n+\delta p_n) = 
I(p_1,\ldots,p_n) + \delta p_1^{\mu}
\frac{\partial}{\partial p_1^{\mu}} 
I(p_1,\ldots,p_n) + \ldots + \delta p_n^{\mu}
\frac{\partial}{\partial p_n^{\mu}} 
I(p_1,\ldots,p_n)\; ,
\end{equation}
one arrives at
\begin{equation}
\delta \epsilon^{\mu}_{\nu} \left( p_1^{\nu}\frac{\partial}{\partial
    p_1^{\mu}} + \ldots +  p_n^{\nu}\frac{\partial}{\partial
    p_n^{\mu}} \right) I(p_1,\ldots,p_n) = 0\; .
\end{equation}
Since $\delta \epsilon^{\mu}_{\nu}$ has six independent components, the 
above equation contains up to six LI identities. These are however not
always linearly independent. To determine the maximum number of linearly 
independent identities, one uses the antisymmetry of $\delta
\epsilon^{\mu}_{\nu}$ to obtain
\begin{equation}
\left(p_1^{\nu}\frac{\partial}{\partial
    p_{1\mu}} - p_1^{\mu}\frac{\partial}{\partial
    p_{1\nu}} + \ldots + p_n^{\nu}\frac{\partial}{\partial
    p_{n\mu}} - p_n^{\mu}\frac{\partial}{\partial
    p_{n\nu}}\right) I(p_1,\ldots,p_n) = 0 \;.
\label{eq:li}
\end{equation}
This equation can be contracted with all possible antisymmetric
combinations of $p_{i\mu}p_{j\nu}$ to yield LI identities for $I$. 
For a three-point vertex, two of the external momenta are linearly
independent ($n=2$), and only one antisymmetric combination of them can
be constructed, resulting in one LI identity ($N_{{\rm LI}}=1$). 
A four-point function
depends on three external momenta ($n=3$), allowing us to construct three
linearly independent antisymmetric combinations, which yield three 
LI identities ($N_{{\rm LI}}=3$). 
The full potential of the LI identities can only be
exploited for integrals involving five or more external legs, which 
allow to construct six linearly independent antisymmetric combinations
of external momenta, thus projecting out all six LI identities 
($N_{{\rm LI}}=6$).

Since $I$ is a scalar, it can not depend on the momenta $p_i$ itself,
but only on scalar products $s_{ij} = 2 p_i\cdot p_j$ of the external
momenta. Replacing 
\begin{equation}
\frac{\partial}{\partial p_{i\mu}} = \sum_j 2\left(p_{i\mu}+p_{j\mu}\right)
\frac{\partial}{\partial s_{ij}}\;,
\end{equation}
one finds that (\ref{eq:li}) becomes a trivial identity, independent of 
$I$. However, the derivatives in (\ref{eq:li}) can be interchanged with
the loop integrations in $I$, such that they do not act anymore on the
integral $I$, but on the integrand of $I$. After this interchange, 
(\ref{eq:li}) becomes a non-trivial relation between different
integrals. 

In the case of two-loop four-point functions, one has a total of 13
equations (10 IBP + 3 LI) for each integrand corresponding to an 
integral of class $I_{t,r,s}$, relating integrals of the same topology
with up to $s+1$ scalar products and $r+1$ denominators, plus integrals
of simpler topologies ({\it i.e.}~with a smaller number of different 
denominators). 
The 13 identities obtained starting from an integral $I_{t,r,s}$ do contain
integrals of the following types:
\begin{itemize}
\item $I_{t,r,s}$: the integral itself. 
\item $I_{t-1,r,s}$: simpler topology. 
\item $I_{t,r+1,s}, I_{t,r+1,s+1}$ : same topology, more complicated than
  $I_{t,r,s}$.  
\item $I_{t,r-1,s}, I_{t,r-1,s-1}$: same topology, simpler than 
   $I_{t,r,s}$.  
\end{itemize}
Quite in general, single identities of the above kind can be used 
to obtain the reduction of $I_{t,r+1,s+1}$ or $I_{t,r+1,s}$ integrals 
in terms of $I_{t,r,s}$ and simpler integrals - rather than to 
get information on the $I_{t,r,s}$ themselves. 

Integrations-by-parts identities are widely applied in multi-loop 
calculations of 
inclusive quantities (see e.g.~\cite{krev} for a review), which are related to 
two-point functions. In these calculations, 
only a relatively small number of different topologies has to be 
considered, but the 
integrals appearing in the calculation can bear large powers 
of propagators and 
scalar products, arising for example from expansions in masses or 
momenta. In these 
calculations, it is desirable to have reduction formulae for 
arbitrary powers of propagators and 
scalar products. These can be obtained from IBP identities derived 
for an integral with arbitrary powers (to be treated symbolically) 
of scalar products and propagators; the derivation 
of these symbolic reduction formulae requires a lot of ingenuity, 
based on the direct inspection of the explicit form of the IBP identities 
for each considered topology, and could not be carried 
out mechanically. 

For loop integrals with a large number of external legs, IBP identities 
are needed for a large number of different topologies, but in general 
for relatively small powers of 
propagators and scalar products. In this case, it would therefore 
be desirable to 
have a mechanical procedure for solving, for any given topology, 
IBP and LI identities for integrals with fixed 
powers of the  propagators and scalar products. 

If one considers the set of all the identities obtained starting from 
the integrand of all the $N(I_{t,r,s})$ integrals of class $I_{t,r,s}$, 
one obtains 
$(N_{{\rm     IBP}}+ N_{{\rm LI}}) N(I_{t,r,s})$ identities 
which contain $N(I_{t,r+1,s+1})+N(I_{t,r+1,s})$ 
integrals of more complicated structure. From (\ref{eq:intnum}) it can
be read off that with increasing $r$ and $s$
the number of identities grows faster than the number
of new unknown integrals\footnote{The importance of this fact was 
first pointed out by S.~Laporta and exploited in~\cite{laporta}.}. 
As a consequence, if for a given $t$-topology one considers the set of 
all the possible equations obtained by considering all the integrands up to 
certain values $r^*, s^*$ of $r, s$, for large enough $r^*, s^*$ 
the resulting system of equations is overconstrained and can be used for 
expressing the more complicated integrals, with greater values of $r, s$ 
in terms of simpler ones, with smaller values of $r, s$. (Let us observe 
that, the system being overconstrained, the equations cannot be all 
independent; it is not {\it a priori} known how many equations are in 
fact linearly independent and, correspondingly, how many integrals 
of the topology under consideration
will remain after reduction).

The required values $r^*$ and $s^*$ for $r$ and $s$ 
can be found by counting the number of accumulated equations (equations
for all integrals with $r\leq r^*$ and $s\leq s^*$) and comparing them
with the number of accumulated unknown integrals, with 
$(r\leq r^* + 1,s \leq s^* + 1)$, but excluding 
$(r=t,s=s^* + 1)$. As an example, we list in Table~\ref{tab:unknown} 
the number of equations and unknowns for two-loop
four-point functions with
seven denominators $t=7$. It can be seen that a complete reduction
requires at least one of the combinations ($r^*,s^*$): (7,2); (8,1);
(9,0).  
\begin{table}[h]
\fbox{$t=7$} \\
\parbox{7.5cm}{\rule[-4mm]{0cm}{1cm}different $I_{t,r,s}$} 
\parbox{7.5cm}{accumulated \parbox{4cm}{equations\\unknowns}}\\
\parbox{7.5cm}{
\begin{tabular}{|r||r|r|r|r|r|}\hline
\backslashbox{$r$}{$s$}
          &    0 &    1 &    2 &    3 &    4 \\ \hline\hline
7 \hspace{0.3cm} &    1 &    2 &    3 &    4 &    5 \\ \hline
8 \hspace{0.3cm} &    7 &   14 &   21 &   28 &   35 \\ \hline
9 \hspace{0.3cm} &   28 &   56 &   84 &  112 &  140 \\ \hline
10 \hspace{0.3cm} &   \phantom{3}84 &  168 &  252 &  336 &  420\\  \hline 
\end{tabular}
}
\parbox{7.5cm}{
\begin{tabular}{|r||r|r|r|r|}\hline
\backslashbox{$r$}{$s$}
  &    0 &    1 &    2 &    3  \\ \hline\hline
  &    13 & 39 & 78 & 130 \\ 
\raisebox{1.5ex}[-1.5ex]{7}\hspace{0.3cm} & 22 & 45 & 76 & 115 \\ \hline
  &   104 & 312 & 624 & 1040 \\
\raisebox{1.5ex}[-1.5ex]{8}\hspace{0.3cm} & 106 & 213 & 354 & 535 \\ \hline
  &   468 & 1404 & 2808 & 4680 \\
\raisebox{1.5ex}[-1.5ex]{9}\hspace{0.3cm} & 358 & 717 & 1196 & 1795 \\ \hline
\end{tabular}
}
\caption{Comparison of the number of (IBP and LI) identities
 to the number of new unknowns (different integrals with 
 $t=7$) appearing in these equations for a two-loop 
 box integral with $t=7$ internal propagators. The identities for
 $I_{t,r,s}$ contain at most $I_{t,r+1,s+1}$. It can be seen that for
 growing $r$ and $s$, the number of equations, upper number in each 
 box, exceeds the number of unknowns, given by the lower number.} 
\label{tab:unknown}
\end{table}

The above table illustrates that typically hundreds of equations have to be
solved in order to obtain a reduction towards simpler integrals. The 
task is performed automatically (and independently of the topology!) 
by a computer program invoking repeatedly the 
computer algebra packages FORM~\cite{form} and MAPLE~\cite{maple}. 
For any given four-point two-loop topology, 
this procedure can result either in a reduction
 towards a small number (typically one or two) of integrals of the
topology under consideration and integrals of simpler 
topology (less different denominators), or even in a complete 
reduction of all integrals of the topology under consideration
towards integrals with simpler topology.
Left-over integrals of the topology under consideration are called 
irreducible master integrals or just 
master integrals. If a topology turns out to contain irreducible master
integrals, one is in principle free to choose which integrals are taken
as master integrals, as long as the chosen integrals are not related by
the IBP and LI identities. In our reduction, we choose 
$I_{t,t,0}$ for 
topologies with one master integral and $I_{t,t,0}$ together 
with the required number of integrals of type 
$I_{t,t+1,0}$ for topologies with more than
one master integral. 

As a final point, it is worth noting that the procedure described above can 
be used to reduce a tensor integral towards scalar integrals. All 
integrals appearing in the projection of an arbitrary tensorial integral 
onto a tensor basis will take the form (\ref{eq:generic}), i.e.~they can 
be classified as $I_{t,r,s}$ and reduced to master integrals.

In the following section, we will demonstrate that the established reduction 
of all the integrals to a few master integrals can be used also to write 
differential equations in the external invariants $s_{ij}$ for the master 
integrals themselves, and then how the differential equations can be used 
to compute these master integrals. 

\section{Differential Equations for Master Integrals}
\label{sec:diffeq}
\setcounter{equation}{0}

The IBP and LI identities discussed in the previous section allow us to
express integrals of the form (\ref{eq:generic}) as a linear
combination of a few master integrals, {\it i.e.} integrals which are 
not further reducible, but have 
to be computed by some different method. 

For the case of massless two-loop 
four-point functions, several techniques have been proposed in the
literature, such as for example the application of a Mellin-Barnes
transformation to all propagators~\cite{smirnov,tausk} 
or the negative dimension
approach~\cite{glover}. Both techniques rely on an explicit integration 
over the loop momenta, with differences mainly in the representation
used for the propagators.
So far, these techniques were only applied to a
limited number of master integrals: Smirnov~\cite{smirnov}
has recently used the
Mellin-Barnes method to compute the planar double box integral
for the case of all external legs on shell (massless case); 
the same method has been applied 
by Tausk~\cite{tausk} for the computation of the non-planar on-shell
double box integral; 
the negative dimension
approach has been applied by Anastasiou, Glover and
Oleari~\cite{glover} to compute the class of two-loop box integrals
which correspond to a one-loop bubble insertion in one of the
propagators of the one-loop box. A general method for the computation of 
the master integrals appearing in two-loop four-point functions 
has however not yet been found. So far, it has also not even 
 been clear 
(apart from the planar double box topology, where Smirnov and Veretin
have recently demonstrated the reducibility of any integral of this
topology to two master
integrals~\cite{smir2}) how many master integrals exist for a given 
topology. Solving the identities discussed in the previous section, we
are now able to identify the irreducible master integrals. A list of
reducible two-loop four-point topologies will be given in
Section~\ref{sec:twoloop}. 

A method for the computation of master integrals avoiding the explicit
integration over the loop momenta is to derive differential equations in 
internal propagator masses or in external momenta for the master integral, 
and to solve these with appropriate boundary conditions. 
This method has first been suggested by Kotikov~\cite{kotikov} to relate 
loop integrals with internal masses to massless loop integrals. 
It has been elaborated in detail and generalised to differential 
equations in external momenta in~\cite{remiddi}; first 
applications were presented in~\cite{appl}.
In the case of four-point functions with one external off-shell leg
and no internal masses, one has three independent
invariants, resulting in three differential equations.

The derivatives in the invariants $s_{ij}=(p_i+p_j)^2$ 
can be expressed by derivatives in the external momenta:
\begin{eqnarray}
\sab \frac{\partial}{\partial \sab} & = & \frac{1}{2} \left( +
p_1^{\mu} \frac{\partial}{\partial p_1^{\mu}} +
p_2^{\mu} \frac{\partial}{\partial p_2^{\mu}} -
p_3^{\mu} \frac{\partial}{\partial p_3^{\mu}}\right) \nonumber \\
\sac \frac{\partial}{\partial \sac} & = & \frac{1}{2} \left( +
p_1^{\mu} \frac{\partial}{\partial p_1^{\mu}} -
p_2^{\mu} \frac{\partial}{\partial p_2^{\mu}} +
p_3^{\mu} \frac{\partial}{\partial p_3^{\mu}}\right)
\label{eq:derivatives} \\
\sbc \frac{\partial}{\partial \sbc} & = & \frac{1}{2} \left( - 
p_1^{\mu} \frac{\partial}{\partial p_1^{\mu}} +
p_2^{\mu} \frac{\partial}{\partial p_2^{\mu}} +
p_3^{\mu} \frac{\partial}{\partial p_3^{\mu}}\right) \nonumber 
\end{eqnarray}
The combinations of derivatives and momenta appearing on the right hand 
side of (\ref{eq:derivatives}) are obviously linearly independent from 
the combinations appearing in the LI identities (\ref{eq:li}) (which 
vanish identically when acting on a function depending on the scalars 
$\sab,\sac,\sbc$). 
The three derivatives of an integral $I_{t,r,s}(\sab,\sac,\sbc,d)$ 
are not linearly independent, but related due to the properties of $I$
under rescaling of all external momenta:
\begin{equation}
I_{t,r,s}(\sab,\sac,\sbc,d) = \lambda^{-\alpha(d,r,s)} 
I_{t,r,s}(\lambda^2\sab,\lambda^2\sac,\lambda^2\sbc,d)\; ,
\end{equation}
where $\alpha(d,r,s)$ is the mass dimension of the integral. For a $m$-loop
integral in $d$ space-time dimensions
with $r$ powers of denominators and $s$ scalar products, one
finds $\alpha(d,r,s) = m d + 2s -2r$. The above equation yields the
rescaling relation 
\begin{equation}
\left[ - \frac{\alpha}{2} + 
\sab \frac{\partial}{\partial\sab} + \sac
  \frac{\partial}{\partial\sac} +  
\sbc \frac{\partial}{\partial\sbc}  \right] I_{t,r,s}(\sab,\sac,\sbc,d)
=0 \; .
\label{eq:rescale}
\end{equation}

Let us now first consider the case of a topology with only one master
integral, which is chosen to be $I_{t,t,0}$, defined as 
\begin{equation}
 I_{t,t,0}(\sij,s_{jk},s_{ki},d) = 
          \int \frac{\d^d k}{(2\pi)^d}\frac{\d^d l}{(2\pi)^d} 
           f_{t,t,0}(k,l,p_i) \; ,
\label{eq:tt0} 
\end{equation} 
where $ f_{t,t,0}(k,l,p_i) $ is a suitable integrand of the form 
appearing in (\ref{eq:generic}). 
It is evident that acting with the right hand  sides of (\ref{eq:derivatives}) 
on the right hand side of (\ref{eq:tt0}) will, after interchange of 
derivative and integration, yield a
a combination of 
integrals of the same type as appearing in the IBP and LI identities for 
$I_{t,t,0}$, including integrals of type $I_{t,t+1,1}$ and 
$I_{t,t+1,0}$. Consequently, the scalar derivatives (on left hand side of 
(\ref{eq:derivatives}))
of  $I_{t,t,0}$ can be expressed by a linear combination of 
integrals up to  $I_{t,t+1,1}$ and 
$I_{t,t+1,0}$.
These can all be reduced to $I_{t,t,0}$ and to integrals of simpler topology
by applying the IBP and LI identities. This reduction results in 
differential equations for $I_{t,t,0}$ of the form:
\begin{eqnarray}
\sij \frac{\partial}{\partial \sij} I_{t,t,0}(\sij,s_{jk},s_{ki},d) & = &
A(\sij,s_{jk},s_{ki},d)
I_{t,t,0}(\sij,s_{jk},s_{ki},d)  \nonumber \\
& & +
F(\sij,s_{jk},s_{ki},d,I_{t-1,r,s}(\sij,s_{jk},s_{ki},d))\;,
\label{eq:master1}
\end{eqnarray}
where $\sij$, $s_{jk}$, $s_{ki}$ are the three invariants, 
$A(\sij,s_{jk},s_{ki},d)$ turns out to be a rational function of the 
invariants and of $d$, 
$F(I_{t-1,r,s})$ is a linear combination (with coefficients
depending on $\sij,s_{jk},s_{ki}$ and $d$) 
of integrals of type 
$I_{t-1,r,s}$, containing only topologies simpler than $I_{t,t,0}$, but
potentially with high powers of some denominators and scalar products. 
These $I_{t-1,r,s}$, which refer to a simpler topology and therefore can be 
considered as known
in a bottom-up approach, play the role of an inhomogeneous term in the 
equation; one can then look for the proper solution
of (\ref{eq:master1}) in a straightforward way. 

The master integral
$I_{t,t,0}(\sij,s_{jk},s_{ki},d)$ can indeed be obtained 
by matching the general
solution of  (\ref{eq:master1}) to an appropriate boundary
condition. Quite in general, finding a boundary condition is 
a  simpler problem than evaluating the whole
integral, since it depends on a smaller number of kinematical
variables. In some cases, the boundary condition can even be
determined from the differential equation itself:
for $\sij=0$, (\ref{eq:master1}) yields, if $A(0,s_{jk},s_{ki},d)\neq 0$, 
\begin{equation}
I_{t,t,0}(0,s_{jk},s_{ki},d) = -
\left[A(0,s_{jk},s_{ki},d)\right]^{-1}  
F(0,s_{jk},s_{ki},d,I_{t-1,r,s}(0,s_{jk},s_{ki},d))\; . 
\end{equation}
For $A$ vanishing at $\sij=0$ one can consider $I_{t,t,0}$ in the limit 
where one of the external momenta vanishes, corresponding to the
vanishing of both invariants involving this momentum, e.g.~$\sij=0$ and 
$s_{ki}=0$ for $p^{\mu}_i=0$.  In this case, 
$I_{t,t,0}$ reduces to a three-point vertex function with one off-shell
external leg. 
All these functions, which might be determined by iterating the procedure 
just described for the considered 4-point function, have actually 
already been computed at the two-loop
level in~\cite{kl} using IBP identities to reduce all possible
topologies to a few master integrals, which, in this case, can be
computed straightforwardly using Feynman parameters
(cf.~Section~\ref{sec:twoloop}). Starting from the
boundary condition in $\sij=s_{ki}=0$, one can determine
$I_{t,t,0}(0,s_{jk},s_{ki},d)$ by solving the differential equation in 
$s_{ki}$ -- this provides the desired boundary condition in
$\sij=0$. 

For topologies with more than one master integral, (\ref{eq:master1})
will be replaced by a system of coupled, linear, first order
differential equations for all master integrals of the topology under
consideration. The determination of the master integrals from these
equations follows the same lines as discussed above, with the only
difference that the general solution for the system of coupled
differential equations is harder to obtain than for a single
equation. In case the coupled equations can not be decoupled by 
an appropriate choice of variables,
several mathematical techniques can be employed
here~\cite{kamke}: the system of $n$ coupled first order 
differential equations can for 
example be rewritten into one $n$-th order differential equation, which
is then solved with standard methods. In some cases, the system can also 
be transformed into a form which is known to be solved by 
generalised hypergeometric series~\cite{kamke,bateman,grad,exton}. 

It is clear from the above discussion, that the determination of a
master integral of a certain topology with $t$ different denominators  
requires the knowledge of all the integrals appearing in the inhomogeneous
term. These integrals are subtopologies of the topology of the 
integral under
consideration, and contain at most $t-1$ different denominators.
The determination of master integrals has therefore to proceed 
bottom-up from
simpler topologies with a small number of different
 denominators towards more complicated
topologies with an increasing number of 
different denominators. For the case under special
consideration, massless 
two-loop four-point functions with up to one external leg 
off-shell, this implies that one has to progress from the simplest 
master integrals with $t=3$ (off-shell two-point function) to construct
all master integrals up to $t=7$. 

\section{A Pedagogical Example: the One-Loop Four-Point Function}
\label{sec:oneloop}
\setcounter{equation}{0}

To illustrate how the method explained above works in practice, we 
present in this section a detailed and self-contained derivation of the 
one-loop four-point function  
\begin{equation}
\boxLO = \int \frac{\d^d k}{(2\pi)^d} \frac{1}{k^2 (k-p_2)^2
  (k-p_2-p_3)^2 (k-p_1-p_2-p_3)^2}\; . 
\label{eq:boxLOdef}
\end{equation}

The topology of this integral is given by the set of its
four propagators, it is the only $t=4$ topology at one loop. In the 
notation introduced above, (\ref{eq:boxLOdef}) corresponds to 
an integral $I_{4,4,0}$, the simplest integral of this topology with 
all propagators appearing in first power and with no scalar products.
For the one-loop four-point function, one has
 four independent scalar products involving $k_{\mu}$ ($k^2$ and
 $k_{\mu}p_i^{\mu}$) and four linear independent denominators. This
 implies that any scalar product involving $k_{\mu}$ can be rewritten as 
 linear combination of propagators and invariants $s_{ij} \equiv 2 p_{i \mu}
 p_j^{\mu}$.
Consequently, 
no integrals with scalar products in the numerator can appear for this  
topology. 

The reduction of all four integrals with one squared propagator ($I_{4,5,0}$)
can be carried out by considering the IBP identities for 
 $I_{4,4,0}$:
\begin{equation}
 \int \frac{\d^d k}{(2\pi)^d} \frac{\partial}{\partial k^{\mu}}
\frac{v^{\mu}}{k^2 (k-p_2)^2
  (k-p_2-p_3)^2 (k-p_1-p_2-p_3)^2}  =  0 \;,
\label{eq:IBPLO}
\end{equation}
where $v^{\mu}$ can be the loop momentum  
$k^{\mu}$ or any of the external momenta
$p_i^{\mu}$, thus yielding four identities. It turns out that 
LI identities and IBP identities for integrals with higher powers of the 
propagators do not contain additional information which would allow 
a reduction of $I_{4,4,0}$. The integral (\ref{eq:boxLOdef}) is therefore 
a master integral. It is the only master integral for this topology. 

To proceed towards the differential equations in the invariants $\sab$, 
$\sac$ and $\sbc$ for this 
master integral, let us consider the derivatives in the external 
momenta:
\begin{eqnarray}
p_1^{\mu} \frac{\partial}{\partial p_1^{\mu}} \boxLO & = &
\int \frac{\d^d k}{(2\pi)^d} \frac{1}{k^2 (k-p_2)^2
  (k-p_2-p_3)^2 (k-p_1-p_2-p_3)^2} \nonumber \\
& & 
\left( \frac{2p_1^{\mu}
    (k-p_1-p_2-p_3)_{\mu}}{(k-p_1-p_2-p_3)^2}\right)\;, \\
 p_2^{\mu} \frac{\partial}{\partial p_2^{\mu}} \boxLO & = &
\int \frac{\d^d k}{(2\pi)^d} \frac{1}{k^2 (k-p_2)^2
  (k-p_2-p_3)^2 (k-p_1-p_2-p_3)^2}\nonumber \\
& & 
 \left(\frac{2p_2^{\mu}
    (k-p_1-p_2-p_3)_{\mu}}{(k-p_1-p_2-p_3)^2} + \frac{2p_2^{\mu}
    (k-p_2-p_3)_{\mu}}{(k-p_2-p_3)^2} + \frac{2p_2^{\mu}
    (k-p_2)_{\mu}}{(k-p_2)^2}\right)\;,  \\
 p_3^{\mu} \frac{\partial}{\partial p_3^{\mu}} \boxLO & = &
\int \frac{\d^d k}{(2\pi)^d} \frac{1}{k^2 (k-p_2)^2
  (k-p_2-p_3)^2 (k-p_1-p_2-p_3)^2} \nonumber \\
& & 
\left(\frac{2p_3^{\mu}
    (k-p_1-p_2-p_3)_{\mu}}{(k-p_1-p_2-p_3)^2} + \frac{2p_3^{\mu}
    (k-p_2-p_3)_{\mu}}{(k-p_2-p_3)^2} \right)\; . 
\end{eqnarray}

 The right hand sides of the above equations contain
 terms with 
\begin{itemize}
\item[{(i)}] four different propagators with one squared propagator, no
  scalar product
\item[{(ii)}] four different propagators, no squared propagator, no
  scalar product
\item[{(iii)}] three different propagators, one squared propagator.
\end{itemize}
The terms of type (i) can now be reduced to type (ii) and (iii) by using 
the integration-by-parts identities (\ref{eq:IBPLO}).
One obtains: 
\begin{eqnarray}
p_1^{\mu} \frac{\partial}{\partial p_1^{\mu}} \boxLO & = & - \boxLO 
+ \boxLOpamb\; , \nonumber \\
p_2^{\mu} \frac{\partial}{\partial p_2^{\mu}} \boxLO & = & - \boxLO 
+ \boxLOpdmc\; , \nonumber \\
p_3^{\mu} \frac{\partial}{\partial p_3^{\mu}} \boxLO & = & (d-6) \boxLO 
- \boxLOpamb - \boxLOpdmc\; , \nonumber 
\end{eqnarray}
where $(\bullet)$ denotes a squared propagator 
and $(\times)$ stands for a pinched (cancelled) propagator. 

A check on these equations is provided by the rescaling
relation (\ref{eq:rescale})
\begin{equation}
\left[ 4-\frac{d}{2} + \sab \frac{\partial}{\partial\sab} + \sac
  \frac{\partial}{\partial\sac} +  
\sbc \frac{\partial}{\partial\sbc}  \right] \boxLO = 0\; ,
\end{equation}
which is related to the above equations by
\begin{equation}
\sab \frac{\partial}{\partial\sab} + \sac
  \frac{\partial}{\partial\sac} +  
\sbc \frac{\partial}{\partial\sbc} = \frac{1}{2} \sum_{i=1}^3 p_i^{\mu}
\frac{\partial}{\partial p_i^{\mu}} \; .
\end{equation}
Inserting the derivatives obtained above, one finds that the rescaling
relation is indeed fulfilled. 

The three-propagator terms appearing in the above equations can be further 
reduced by using the IBP identities for the corresponding vertex; 
the vertex amplitudes, in turn, can all be expressed in terms of bubble 
integrals, and one finally finds 
\begin{equation}
\boxLOpamb = \frac{d-3}{p_2\cdot (p_1+p_3)} \left[
  \frac{1}{(p_1+p_2+p_3)^2} \bubbleLO{p_{123}} - \frac{1}{(p_1+p_3)^2}
  \bubbleLO{p_{13}} \right]\; .
\end{equation}
A similar identity is obtained by exchanging $p_1
\leftrightarrow p_2$. 

The differential equations for the one-loop box integral then follow 
from (\ref{eq:derivatives}). 
The set of differential equations reads:
\begin{eqnarray}
\sab \frac{\partial}{\partial \sab} 
\boxLO
& = & - \frac{d-4}{2} \boxLO \nonumber \\
& & + \frac{2(d-3)}{\sab+\sac}\left[\frac{1}{\sabc} \bubbleLO{p_{123}}
  -\frac{1}{\sbc} \bubbleLO{p_{23}} \right] \nonumber \\
& & + \frac{2(d-3)}{\sab+\sbc}\left[\frac{1}{\sabc} \bubbleLO{p_{123}}
  -\frac{1}{\sac} \bubbleLO{p_{13}} \right]\;, \label{eq:boxLOsab}\\
\sac \frac{\partial}{\partial \sac} 
\boxLO
& = &  \frac{d-6}{2} \boxLO \nonumber \\
& & - \frac{2(d-3)}{\sab+\sac}\left[\frac{1}{\sabc} \bubbleLO{p_{123}}
  -\frac{1}{\sbc} \bubbleLO{p_{23}} \right]\;, \label{eq:boxLOsac}\\
\sbc \frac{\partial}{\partial \sbc} 
\boxLO
& = &  \frac{d-6}{2} \boxLO \nonumber \\
& & - \frac{2(d-3)}{\sab+\sbc}\left[\frac{1}{\sabc} \bubbleLO{p_{123}}
  -\frac{1}{\sac} \bubbleLO{p_{13}} \right]\;, \label{eq:boxLOsbc}
\end{eqnarray}
where $\sabc = \sab+\sac + \sbc$. The one-loop bubble diagrams in the
inhomogenous term yield:
\begin{equation}
\bubbleLO{p} = 
\left[\frac{(4\pi)^{\frac{4-d}{2}}}{16\pi^2}\frac{ \Gamma (3-d/2)
    \Gamma^2 (d/2-1)}{ \Gamma (d-3)} \right]  \; \frac{-2i}{(d-4)(d-3)}
\left( -p^2\right)^{\frac{d-4}{2}} \equiv A_{2,LO} 
\left( -p^2\right)^{\frac{d-4}{2}} \;.
\end{equation}

The boundary conditions in $\sij=0$ can be readily read off from the
above:
\begin{eqnarray}
\boxLO(\sab=0) & = & \frac{4(d-3)}{(d-4)}\frac{1}{\sac\sbc}\Bigg[
  \bubbleLO{p_{123}} \nonumber \\
& & \hspace{1.3cm}
- \bubbleLO{p_{13}} -
  \bubbleLO{p_{23}}  \Bigg]\;, \label{eq:boundLOsab} \\ 
\boxLO(\sac=0) & = & \frac{4(d-3)}{(d-6)}\frac{1}{\sab}\Bigg[\frac{1}{\sabc}
  \bubbleLO{p_{123}} - \frac{1}{\sbc} \bubbleLO{p_{23}}  \Bigg]\;, \\ 
\boxLO(\sbc=0) & = & \frac{4(d-3)}{(d-6)}\frac{1}{\sab}\Bigg[\frac{1}{\sabc}
  \bubbleLO{p_{123}} - \frac{1}{\sac} \bubbleLO{p_{13}}  \Bigg] \;.
\end{eqnarray}
The result for the one-loop box integral can in principle 
be obtained by integrating 
any of the differential equations
(\ref{eq:boxLOsab})--(\ref{eq:boxLOsbc}). In practice,
it turns out to be more appropriate to introduce a new set of 
variables, namely $\sac$, $\sbc$ and $\sabc=\sab+\sac+\sbc$, corresponding
to the arguments appearing in the two-point functions in the
inhomogeneous terms. This transformation yields a differential equation
in $\sabc$, which will be used for integration. Note that this
transformation also modifies the differential equations in $\sac$ and
$\sbc$. The differential equation in $\sabc$ reads:
\begin{eqnarray}
\lefteqn{\frac{\partial}{\partial \sabc} 
\boxLO
+  \frac{d-4}{2(\sabc-\sac-\sbc)} \boxLO = } \nonumber \\
& & + \frac{2(d-3)}{(\sabc-\sbc)(\sabc-\sac-\sbc)}
\left[\frac{1}{\sabc} \bubbleLO{p_{123}}
  -\frac{1}{\sbc} \bubbleLO{p_{23}} \right] \nonumber \\
& & + \frac{2(d-3)}{(\sabc-\sac)(\sabc-\sac-\sbc)}
\left[\frac{1}{\sabc} \bubbleLO{p_{123}}
  -\frac{1}{\sac} \bubbleLO{p_{13}} \right] \label{eq:boxLOq}\; .
\label{eq:boxLOmaster}
\end{eqnarray}
The boundary condition in $\sabc=0$ can not be trivially determined from
this equation, reflecting the fact that the massless one-loop 
box integral with all external legs on shell is not reducible to simpler 
subtopologies by IBP identities. The boundary condition in
$\sabc=-\sac-\sbc$ can however be determined from (\ref{eq:boundLOsab}). 

Equation~(\ref{eq:boxLOmaster}) is a linear, inhomogeneous first order 
differential equation of the form
\begin{displaymath}
\frac{\partial y(x)}{\partial x} + f(x) y(x) = g(x),
\end{displaymath}
which can be solved by 
introducing an integrating factor (see for instance~\cite{bronstein} or 
any standard book on differential equations) 
\begin{displaymath}
M(x) = e^{\int f(x) \d x},  
\end{displaymath}
such that  $y(x)=1/M(x)$ solves the homogenous differential equation
($g(x)=0$). This 
yields the general solution of the inhomogenous equation as
\begin{displaymath}
y(x) = \frac{1}{M(x)} \left( \int g(x) M(x) \d x + C\right),
\end{displaymath}
where the integration constant $C$ can be adjusted to match the boundary 
conditions. 

For~(\ref{eq:boxLOmaster}), we have at once the integrating factor
\begin{equation}
M(\sabc) = (\sac + \sbc - \sabc )^{\frac{d-4}{2}}\; . 
\label{eq:intfac}
\end{equation}
This factor is not unambiguous, since 
\begin{equation}
M'(\sabc) = (\sabc -\sac - \sbc)^{\frac{d-4}{2}}\; . 
\label{eq:intfacal}
\end{equation}
would also be a valid integrating factor. We select 
(\ref{eq:intfac}) by requiring a real integrating factor 
in the region:
$-\sabc \ge -\sac - \sbc \ge 0$. The final result for the 
box integral does
not depend on the selection of the integrating factor; using
(\ref{eq:intfacal}), one must however be more careful in applying
analytic continuation formulae and in multiplying non-integer powers of
the invariants. 

With the integrating factor (\ref{eq:intfac}), the one-loop box integral 
reads:
\begin{eqnarray}
\boxLO (\sabc,\sac,\sbc) &=& 
2(d-3) A_{2,LO} \left( \sac + \sbc - \sabc
\right)^{2-\frac{d}{2}}
\int^{\sabc}\hspace{-0.25cm} \d \sabcp \left( \sac + \sbc - \sabcp
\right)^{\frac{d}{2}-3}\nonumber \\
& & \hspace{-0.9cm}
\Bigg[  \frac{\left(-\sac\right)^{\frac{d}{2}-3}}{ \sac - \sabcp }
       +\frac{\left(-\sbc\right)^{\frac{d}{2}-3}}{\sbc - \sabcp}
 - \frac{2\sabcp - \sac - \sbc}{(\sac - \sabcp)(\sbc - \sabcp) } 
    \left(-\sabcp\right)^{\frac{d}{2}-3} \Bigg]. 
\label{eq:boxLOintform}
\end{eqnarray}
From a computational point of view, the evaluation of the box amplitude 
of (\ref{eq:boxLOdef}) has been reduced to the one dimensional integration 
corresponding to the solution of (\ref{eq:boxLOmaster}). 

The lower boundary of the integral is independent of $\sabc$ and can be
adjusted arbitrarily. The first two terms in (\ref{eq:boxLOintform}) 
can be easily integrated by shifting the integration variable to
$\sabcp - \sac - \sbc$, which is then integrated between $0$ and 
 $\sabc - \sac - \sbc$. To integrate the last term, one introduces a new
 variable $\sabcp (\sabcp - \sac - \sbc)$, which is integrated between 
$0$ and $\sabc (\sabc - \sac - \sbc)$. The resulting integrals yield 
can be identified as integral representation of the hypergeometric function
$\,_2F_1$.
With this choice of variables and
boundaries, no constant term is required to match the boundary
conditions. The result for the one-loop box integral then reads:
\begin{eqnarray}
\lefteqn{\boxLO   =  - \frac{4(d-3)}{d-4} A_{2,LO}
\frac{1}{\sac\sbc}} \nonumber \\ 
 & & \hspace{0.46cm}
\Bigg[ \left(\frac{\sac \sbc}{\sac - \sabc}\right)^{\frac{d}{2}-2}
 \,_2F_1\left(d/2-2, d/2-2; d/2-1; \frac{\sabc - \sac - \sbc}{\sabc -\sac}
 \right)\nonumber \\ 
 & & \hspace{0.4cm}
+ \left(\frac{\sac \sbc}{\sbc - \sabc}\right)^{\frac{d}{2}-2}
 \,_2F_1\left(d/2-2, d/2-2; d/2-1; \frac{\sabc - \sac - \sbc}{\sabc -\sbc}
 \right) \nonumber \\
& & \hspace{0.4cm} - \left( \frac{-\sabc \sac \sbc}{(\sac - \sabc)
   (\sbc - \sabc)} 
  \right)^{\frac{d}{2}-2} \,_2F_1\left(d/2-2, d/2-2; d/2-1; 
\frac{\sabc (\sabc - \sac - \sbc)}{(\sabc -\sac)(\sabc - \sbc)}\right) \Bigg]
\;.
\nonumber \\
\end{eqnarray}
The invariants appearing in this expression
can be safely continued from the region $-\sabc \ge -\sac - \sbc \ge 0$ 
to the physical  region $\sabc \ge  \sac + \sbc \ge 0$. 
The arguments of the hypergeometric functions are
ratios of invariants, they are not changed by the analytic
continuation. The non-integer powers of invariants appearing as
coefficients acquire imaginary parts, their signs are uniquely
determined by the convention $-p^2 = -p^2 - i0$,
thus fixing the imaginary part of the whole expression. 
The above equation
reproduces the well-known result from the literature, e.g.~\cite{ert}. 
It should be kept in mind that in applying dimensional regularisation, no 
distinction between infrared and ultraviolet poles, which both 
show up as 
$1/(d-4)$ in the above equation and in the coefficient $A_{2,LO}$, is
made.
 Needless to say, those 
singularities are an intrinsic feature of the dimensional regularisation, 
and by non means an artifact of the differential equation approach. 
In particular, the most singular part of the above one loop integral,
which is ultraviolet finite,  as 
well as of the two loop integrals discussed in the following section,
arises from soft configurations, and can in principle be re-derived by 
applying a strong ordering procedure~\cite{catani} to the integrands.

When the box integral is expressed in the above form, 
where no expansion around $d=4$ has yet
been performed, analytic continuations, e.g.~to the on-shell case $\sabc=0$ or
to collinear and soft limits $s_{ij}=0$, can be made with ease.     

\section{Results on Two-Loop Four-Point Functions}
\label{sec:twoloop}
\setcounter{equation}{0}

In the following, we shall outline how the techniques derived in 
Sections \ref{sec:master} and \ref{sec:diffeq} 
can be applied to the computation of two-loop integrals 
appearing in amplitudes for the decay of one massive into 
three massless particles: two-loop four-point functions with one
off-shell leg. The main purpose of this section is to illustrate 
applications of the tools developed above to non-trivial problems; 
the list of integrals given here is far from complete. We provide a 
comprehensive list of master integrals only up to $t=5$, for 
$t=6$ and $t=7$ only reducible integrals are quoted.  

A prefactor common to all massless scalar integrals is 
\begin{equation}
S_d = \left[(4\pi)^{\frac{4-d}{2}}\frac{ \Gamma (3-d/2)
    \Gamma^2 (d/2-1)}{ \Gamma (d-3)} \right] \; ,
\end{equation}
which is also appearing in all counterterms in the $\overline{{\rm
    MS}}$--scheme. 

In the following, the notation of external momenta is as follows:
$p_i$ denotes an on-shell momentum, $p_{ij(k)}$ denotes an off-shell
momentum, being the sum of two (three) on-shell momenta $p_i$, $p_j$(,
$p_k$) with $s_{ij(k)}=(p_{ij(k)})^2$. $p$ is an arbitrary momentum.

\subsection{$t=3$}
For two-loop integrals with $t=3$, only one topology exists: the 
two-loop vacuum bubble. The
corresponding integral fulfils a homogeneous differential equation,
which can not be used to infer any boundary condition. The integral can
however be computed using Feynman parameters:
\begin{eqnarray}
\bubbleNLO{p} &=&  \left(\frac{S_d}{16 \pi^2}\right)^2
  \frac{\Gamma(5-d)\Gamma^2(d-3)}{\Gamma^2(3-d/2)\Gamma(d/2-1)\Gamma(3d/2-3)} 
\frac{1}{(d-3)(d-4)}
 \left( -p^2 \right)^{d-3} \nonumber \\
&\equiv & A_3   \left( -p^2 \right)^{d-3}\;.
\end{eqnarray}

\subsection{$t=4$}
Several different two-loop topologies exist for $t=4$. Two types of
two-point functions are encountered. The first can be reduced to
\begin{equation}
\bubblexNLO{p}= \frac{3d-8}{d-4} \frac{1}{p^2} \bubbleNLO{p}
\end{equation}
using  IBP identities. The  second is the product of two one-loop 
bubble integrals and yields
\begin{equation}
\doublebubbleNLO{p} = \left( A_{2,LO} \right)^2 (-p^2)^{d-4}\; ,
\end{equation}
which generalises trivially to a three-point function
\begin{equation}
\doublebubblexNLO{p_{123}}{p_{12}}{p_3} 
= \left( A_{2,LO} \right)^2 
(-\sabc)^{\frac{d}{2}-2}(-\sab)^{\frac{d}{2}-2}\; .
\label{eq:doublebubblex}
\end{equation}

Only one of the master integrals at $t=4$ fulfils a homogeneous
differential equation: a two-loop vertex integral with one
off-shell leg. This integral can also be computed using Feynman parameters:
\begin{eqnarray}
\triangleNLO{p_{12}}{p_1}{p_2} 
&=& \left(\frac{S_d}{16 \pi^2}\right)^2 \frac{ \Gamma^3(d-3) 
 \Gamma(5-d)}{\Gamma(3-d/2)\Gamma^2(d/2-1)
\Gamma(3d/2-4)}\frac{-2}{(d-3)(d-4)^2} 
\left(-s_{12}\right)^{d-4} \nonumber\\
&\equiv &  A_4 \left(-s_{12}\right)^{d-4}\; .
\end{eqnarray}
The other vertex integral topology with one off-shell leg can be reduced 
using IBP identities:
\begin{equation} 
\trianglexNLO{p_{12}}{p_1}{p_2} = -\frac{3d-8}{d-4} \frac{1}{s_{12}}
\bubbleNLO{p_{12}} \; .
\end{equation}

Among the three vertex integrals with two off-shell legs, only one can
be reduced using IBP and LI identities:
\begin{equation}
\trianglexNLO{p_{123}}{p_{12}}{p_3} = \frac{3d-8}{d-4} \frac{1}{s_{123}
  - s_{12}} \left[ \bubbleNLO{p_{12}} - \bubbleNLO{p_{123}} \right] \;.
\end{equation}
The two remaining ones are master integrals. Written out in terms of 
propagators,  they read:
\begin{eqnarray}
\triangleNLO{p_{123}}{p_{12}}{p_3} & =& 
\int \frac{\d^d k}{(2\pi)^d}\frac{\d^d l}{(2\pi)^d}  \frac{1}{
k^2 l^2 (k-p_{123})^2 (k-l-p_{12})^2} \; , 
\label{eq:firstMI} \\
\trianglexNLO{p_{123}}{p_3}{p_{12}} & =& 
\int \frac{\d^d k}{(2\pi)^d}\frac{\d^d l}{(2\pi)^d}  \frac{1}{
k^2 l^2 (l-p_{12})^2 (k-l-p_{3})^2} \; . 
\end{eqnarray}
Both  fulfil inhomogeneous
differential equations. For a vertex $p_{123} \to p_{12} + p_3$, the
appropriate variables for the differential equations are $s_{123}$ and
$s_{12}$. To illustrate the structure of the differential equations, we
quote them for (\ref{eq:firstMI}):
\begin{eqnarray}
s_{123} \frac{\partial}{\partial s_{123}}
\triangleNLO{p_{123}}{p_{12}}{p_3} & = & \frac{d-4}{2}\; \frac{2\sabc -
  \sab}{\sabc - \sab}  \triangleNLO{p_{123}}{p_{12}}{p_3} \nonumber \\
& &-
\frac{3d-8}{2}\; \frac{1}{\sabc - \sab} \bubbleNLO{p_{12}}\;, \nonumber \\
\sab \frac{\partial}{\partial \sab} \triangleNLO{p_{123}}{p_{12}}{p_3} 
& = & - \frac{d-4}{2}\; \frac{\sab}{\sabc - \sab}
\triangleNLO{p_{123}}{p_{12}}{p_3}  \nonumber \\
&& + 
\frac{3d-8}{2}\; \frac{1}{\sabc - \sab} \bubbleNLO{p_{12}}\;.
\end{eqnarray}
The boundary conditions for $\sabc=0$ or $\sab=0$ are obtained directly
from the vertex integrals with one off-shell leg quoted above. Using
these, one finds 
\begin{equation}
\!\triangleNLO{p_{123}}{p_{12}}{p_3}\! = A_4 \left( \sab -
  \sabc\right)^{\frac{d}{2}-2} \left(-\sabc\right)^{\frac{d}{2}-2}
- \frac{3d-8}{2(d-3)} A_3 \frac{\left(-\sab\right)^{d-3}}{-\sabc}
\,_2F_1\left(\frac{d}{2}-1, 1; d-2; \frac{\sab}{\sabc}\right).
\end{equation}
The second master integral can be obtained from this by analytic
continuation of the hypergeometric function:
\begin{equation}
\trianglexNLO{p_{123}}{p_3}{p_{12}} = -\frac{3d-8}{d-4} A_3 
\left( -\sab \right)^{\frac{d}{2}-2} \left(-\sabc\right)^{\frac{d}{2}-2}
\,_2F_1\left(\frac{d}{2}-1, 2-\frac{d}{2}; 3-\frac{d}{2}; 
\frac{\sabc-\sab}{\sabc}\right)\; .
\end{equation}

Vertex integrals with three off-shell legs can not appear as
subtopologies in two-loop four-point functions with one off-shell leg.

\subsection{$t=5$}
The two-loop two-point function with $t=5$ is a  well known
example~\cite{chet,krev} for the application of IBP identities: 
\begin{equation}
\bubblecrossNLO{p}  =   
\frac{2(3d-8)(3d-10)}{(d-4)^2} \frac{1}{\left(p^2\right)^2}
\bubbleNLO{p}- \frac{2(d-3)}{d-4} \frac{1}{p^2} \doublebubbleNLO{p}\; .
\end{equation}

The four different $t=5$
three-point functions with one off-shell leg can also
be reduced by using IBP and LI identities:
\begin{eqnarray}
\triangleaNLO{p_{12}}{p_1}{p_2} &= & \frac{(d-3)(3d-8)(3d-10)}{(d-4)^3}
\;\frac{1}{(\sab)^2} \bubbleNLO{p_{12}}\;,\\
\trianglebNLO{p_{12}}{p_1}{p_2} &= & \frac{(d-3)(3d-10)}{(d-4)^2}
\; \frac{1}{\sab} \triangleNLO{p_{12}}{p_1}{p_2} \nonumber \\
&& -\frac{(d-3)(3d-8)(3d-10)}{(d-4)^3}
\;\frac{1}{(\sab)^2} \bubbleNLO{p_{12}}\;,  \\
\trianglecNLO{p_{12}}{p_1}{p_2} &= & -\frac{(3d-8)(3d-10)}{(d-4)^2}
\;\frac{1}{(\sab)^2} \bubbleNLO{p_{12}}\;,\\
\triangledNLO{p_{12}}{p_1}{p_2} &= & -\frac{3d-10}{2(d-4)}
\; \frac{1}{\sab} \triangleNLO{p_{12}}{p_1}{p_2}\;.
\end{eqnarray}

By applying IBP and LI identities, it is likewise possible to reduce
all but one $t=5$ three-point function with two off-shell legs:
\begin{eqnarray}
\triangleaNLO{p_{123}}{p_{12}}{p_3} &= & - \frac{(d-3)(3d-10)}{(d-4)^2}
\; \frac{1}{\sabc - \sab} \trianglexNLO{p_{123}}{p_3}{p_{12}} \nonumber\\
&& + \frac{(d-3)(3d-8)(3d-10)}{(d-4)^3}
\;\frac{1}{\sab(\sabc-\sab)} \bubbleNLO{p_{12}}\;, \\
\trianglebNLO{p_{123}}{p_3}{p_{12}} &= & 
\frac{(d-3)(3d-10)}{(d-4)^2}
\; \frac{1}{\sabc - \sab} \triangleNLO{p_{123}}{p_{12}}{p_3} \nonumber\\
&& - \frac{(d-3)(3d-8)(3d-10)}{(d-4)^3}
\;\frac{1}{\sabc(\sabc-\sab)} \bubbleNLO{p_{123}}\;, \\
\trianglecNLO{p_{123}}{p_{12}}{p_3} &= & \frac{(3d-8)(3d-10)}{(d-4)^2}
\; \frac{1}{\sab(\sabc-\sab)} \bubbleNLO{p_{12}}
\nonumber \\
&& - \frac{(3d-8)(3d-10)}{(d-4)^2}\;\frac{1}{\sabc(\sabc-\sab)}
\bubbleNLO{p_{123}}\;, \\ 
 \trianglecNLO{p_{123}}{p_3}{p_{12}} &= & \frac{3d-10}{2(d-4)}
\; \frac{1}{\sabc-\sab} \trianglexNLO{p_{123}}{p_3}{p_{12}} \nonumber\\
&&  - \frac{(3d-8)(3d-10)}{2(d-4)^2}\;\frac{1}{\sabc(\sabc-\sab)}
\bubbleNLO{p_{123}}\;, \\ 
\triangledNLO{p_{123}}{p_{12}}{p_3} &= & - \frac{3d-10}{2(d-4)}
\; \frac{1}{\sabc-\sab} \triangleNLO{p_{123}}{p_{12}}{p_3}\nonumber\\
&&+ \frac{(3d-8)(3d-10)}{2(d-4)^2}\;\frac{1}{\sab(\sabc-\sab)}
\bubbleNLO{p_{12}}\; .
\end{eqnarray}

The remaining three-point function is a master integral, which can be
found by solving the corresponding differential equations:
\begin{eqnarray}
\trianglebNLO{p_{123}}{p_{12}}{p_3} &= &\frac{2(d-3)}{d-4}\, 
\left(A_{2,LO}\right)^2\;
\left(-\sab\right)^{d-4} \frac{1}{-\sabc}
\,_2F_1\left(
1-\frac{d}{2},d-3; d-2; \frac{\sabc-\sab}{\sabc} \right) \nonumber \\
& & - \frac{(3d-8)(3d-10)}{(d-4)^2} \, A_3 \nonumber \\
& & 
\Bigg[
\frac{1}{\sab}\, \left(-\sabc\right)^{d-4} 
\,_3F_2\left( \frac{d}{2}-1,1,d-3; 3-\frac{d}{2},d-2;
  \frac{\sab-\sabc}{\sab} \right) \nonumber \\
&&  + \frac{1}{\sabc}\, \left(-\sab\right)^{d-4} 
\,_3F_2\left( \frac{d}{2}-1,1,d-3; 3-\frac{d}{2},d-2;
  \frac{\sabc-\sab}{\sabc} \right)\Bigg] \; .
\end{eqnarray}

For $t=5$, one finds four different topologies for two-loop four-point
functions. These are all master integrals obeying inhomogeneous
differential equations in the external invariants. Solving these
equations and matching the boundary conditions, all master integrals 
can be determined. 
\begin{eqnarray}
\boxbubbleaNLO{p_{123}}{p_1}{p_2}{p_3} & = & \frac{3d-10}{d-4}
\, A_4 \, \frac{1}{\sbc}\left(-\sac\right)^{d-4}
\left(\frac{\sab+\sac}{\sbc}\right)^{3-d}\nonumber \\
&& \hspace{1.4cm}
\,_2F_1\left(d-3,\frac{d}{2}-2;\frac{d}{2}-1;\frac{\sab}{\sab+\sac}\right)
\nonumber\\
&& -\frac{3d-10}{2(d-4)}\, A_4 \, \frac{1}{\sab}
\left( -\sab-\sbc\right)^{d-4} 
\left(\frac{\sab+\sac}{\sab}\right)^{\frac{d}{2}-3}\nonumber \\
&& \hspace{1.4cm}
\,_2F_1\left(3-\frac{d}{2},4-d;5-d;\frac{\sac\sbc}{(\sab+\sac)(\sab+\sbc)}
\right)\nonumber \\
&&\hspace{-1cm}
-\frac{(3d-8)(3d-10)}{4(d-3)(d-4)} \,A_3\,
\frac{1}{\sab\sabc}
\left(-\sac\right)^{d-3} 
\left(\frac{\sab+\sbc}{\sabc}\right)^{\frac{d}{2}-2} 
\left(\frac{\sab+\sac}{\sab}\right)^{\frac{d}{2}-3} \nonumber\\
&& \hspace{-1cm}
\,_2F_1\left(\frac{d}{2}-1,1;d-2;\frac{\sac}{\sabc}\right)
\,_2F_1\left(3-\frac{d}{2},4-d;5-d;\frac{\sac\sbc}{(\sab+\sac)(\sab+\sbc) 
   }\right)\nonumber \\
&&\hspace{-1cm}-\frac{(3d-8)(3d-10)}{2(d-4)^2}\, A_3 \, \frac{1}{-\sab-\sbc}
\left(-\sac\right)^{d-4}
\left(\frac{\sab+\sbc}{\sab+\sac}\right)^{\frac{d}{2}-1} 
\left(\frac{\sabc}{\sab}\right)^{\frac{d}{2}-2}\nonumber \\
&& F_1\left(4-d,2-\frac{d}{2},2-\frac{d}{2},5-d,\frac{\sbc}{\sab+\sbc},
\frac{\sbc}{\sabc}\right) \;, \label{eq:firstFIVE}
\\
\boxbubblebNLO{p_{123}}{p_1}{p_2}{p_3} & = & 
\frac{(3d-10)(3d-8)}{(d-4)(d-6)}\, A_3\, \frac{1}{\sab+\sbc} \bigg[
\left(-\sac\right)^{d-4}
\,_2F_1\left(1,1;4-\frac{d}{2};\frac{\sbc}{\sab+\sbc} \right) \nonumber \\
&& \hspace{0.5cm} - \left(-\sabc\right)^{d-4}
F_1\left(1,1,4-d,4-\frac{d}{2};\frac{\sbc}{\sab+\sbc},
\frac{\sbc}{\sabc}\right) \bigg] \nonumber\\ 
&& - \frac{2(3d-8)(3d-10)}{(d-4)(d-6)}\,  A_3\, \left(-\sabc\right)^{d-5}
 \,_2F_1\left(1,2-\frac{d}{2};\frac{d}{2}-1;\frac{\sab}{\sab+\sac}\right)
\nonumber \\
&& \hspace{-0.7cm} \left[ \,_2F_1\left(1,5-d;4-\frac{d}{2}; \frac{\sbc}{\sabc}\right) +
\frac{d-6}{d-4} 
\,_2F_1\left(1,5-d;3-\frac{d}{2}; \frac{\sab+\sac}{\sabc}\right) \right]
\!\! , \label{eq:secondFIVE}
\\
\boxxaNLO{p_{123}}{p_1}{p_2}{p_3} & = & \frac{(3d-8)(3d-10)}{(d-4)^2}\;
A_3 \;\frac{1}{\sac\sbc} \;\Bigg[ \nonumber \\
&& \hspace{-2cm} -\left(\frac{\sac\sbc}{-(\sab+\sbc)}\right)^{d-3}
\,_2F_1\left( d-3,d-3;d-2;\frac{\sab}{\sab+\sbc}\right) \nonumber \\
&& \hspace{-2cm}-\left(\frac{\sac\sbc}{-(\sab+\sac)}\right)^{d-3}
\,_2F_1\left( d-3,d-3;d-2;\frac{\sab}{\sab+\sac}\right) 
\label{eq:thirdFIVE}\\
&&\hspace{-2cm}
 +\left(\frac{-\sabc\sac\sbc}{(\sab+\sbc)(\sab+\sac)}\right)^{d-3}
\,_2F_1\left(
  d-3,d-3;d-2;\frac{\sab\sabc}{(\sab+\sac)(\sab+\sbc)}\right)\Bigg] \;,
\nonumber\\
\boxxbNLO{p_{123}}{p_1}{p_2}{p_3} & = & - A_4 \frac{(d-3)(3d-10)}{(d-4)^2}
\left( -\sabc \right)^{d-4} \left(-\sac-\sbc\right)^{3-d} \nonumber \\
&& \Bigg[ \left(-\sbc\right)^{d-4} F_1\left(4-d,d-3,2-\frac{d}{2},5-d,
\frac{\sac}{\sac+\sbc},\frac{\sac}{\sabc}\right)\nonumber \\
&& +\left(-\sac\right)^{d-4} F_1\left(4-d,d-3,2-\frac{d}{2},5-d,
\frac{\sbc}{\sac+\sbc},\frac{\sbc}{\sabc}\right)\Bigg]\nonumber \\
&& -A_3\frac{(3d-8)(3d-10)}{2(d-4)^2} \frac{1}{-\sabc} \nonumber \\
&& \Bigg[ \left(-\sac\right)^{d-4} S_1\left(\frac{d}{2}-1,1,1,d-2,5-d,
\frac{\sbc}{\sabc},\frac{\sac}{\sabc}\right)\nonumber \\
&& 
+\left(-\sbc\right)^{d-4} S_1\left(\frac{d}{2}-1,1,1,d-2,5-d,
\frac{\sac}{\sabc},\frac{\sbc}{\sabc}\right)\Bigg]\;. \label{eq:fourthFIVE}
\end{eqnarray}
The integrals (\ref{eq:firstFIVE}) and (\ref{eq:secondFIVE})
are one-loop bubble insertions into the one-loop 
box and have already been computed for arbitrary powers of the propagators
in~\cite{glover}. The integrals (\ref{eq:thirdFIVE}) and 
(\ref{eq:fourthFIVE})
were, to our knowledge, not  
known up to now. 
In the reduction of integrals of the topology (\ref{eq:fourthFIVE}), 
one finds two master 
integrals, whose differential equations decouple in the variable 
$\Delta\equiv\sac-\sbc$. The second master integral for this topology
can be found 
by rearranging one of the differential equations:
\begin{eqnarray}
\lefteqn{\boxxbdotNLO{p_{123}}{p_1}{p_2}{p_3} =}\nonumber \\
&&  - \frac{4\sab\sabc}{\sac\sbc}
\frac{\partial}{\partial\sab} \boxxbNLO{p_{123}}{p_1}{p_2}{p_3}\nonumber \\
&& + (d-4)\frac{3\sab+\sabc}{\sac\sbc} \boxxbNLO{p_{123}}{p_1}{p_2}{p_3}
\nonumber \\
&& -\frac{(d-3)(3d-10)}{d-4}\frac{1}{\sac\sbc} \left( 
\triangleNLO{p_{123}}{p_{13}}{p_2} + \triangleNLO{p_{123}}{p_{23}}{p_1}\right)
\nonumber \\
&& +\frac{2(d-3)(3d-8)(3d-10)}{(d-4)^2}\frac{1}{\sac\sbc}\left(
\frac{1}{\sac} \bubbleNLO{p_{13}} + \frac{1}{\sbc}\bubbleNLO{p_{23}} 
\right)\; .
\end{eqnarray}

Finally, products of one-loop vertex with one-loop bubble integrals also 
yield topologies with $t=5$. These can all be reduced to
(\ref{eq:doublebubblex}) and are not quoted explicitly. 

The complete
list of integrals at $t=5$ which were derived in this section can now be 
used to compute all integrals at $t=6$ and $t=7$ which can be reduced
using IBP and LI identities. The results of this reduction are
summarised in the following.

\subsection{$t=6$}
Two-loop integrals with $t=6$ arising in calculations in covariant gauges
must be three- or four-point functions. 
Since we are concerned with subgraphs that can appear in the reduction of 
four-point functions with one off-shell leg, we need to consider 
three-point functions with up to two off-shell legs. For general
three-point functions at $t=6$, one finds three distinct topologies:
two planar and one crossed arrangement of the loop momenta. The crossed graphs 
correspond to master integrals, while the planar graphs are reducible,
as first pointed out in~\cite{kl}, where the three-point integral
with one off-shell leg was computed. We reproduce these results:
\begin{eqnarray}
\triaplanNLO{p_{12}}{p_1}{p_2} &= & \frac{3(d-3)(3d-10)}{(d-4)^2} 
\frac{1}{\sab^2} \triangleNLO{p_{12}}{p_1}{p_2} \nonumber \\
&& + \frac{4(d-3)^2}{(d-4)^2} \frac{1}{\sab^2} \doublebubbleNLO{p_{12}} 
\nonumber \\
&&-\frac{6(d-3)(3d-8)(3d-10)}{(d-4)^3} \frac{1}{\sab^3} \bubbleNLO{p_{12}}
\; ,\\
\triaplanxNLO{p_{12}}{p_1}{p_2} &= & -\frac{3(d-3)(3d-10)}{2(d-4)(d-5)}
\frac{1}{\sab^2} \triangleNLO{p_{12}}{p_1}{p_2} \nonumber \\
&&+ \frac{3(d-3)(3d-8)(3d-10)}{(d-4)^2(d-5)} 
\frac{1}{\sab^3} \bubbleNLO{p_{12}}\; .
\end{eqnarray}
The results for two off-shell legs read:
\begin{eqnarray}
\triaplanNLO{p_{123}}{p_1}{p_{23}} &= & \frac{1}{\sabc} 
\trianglebNLO{p_{123}}{p_{23}}{p_1} + \frac{(d-3)(3d-10)}{(d-4)^2}
\frac{1}{(\sab+\sac)\sabc} \trianglexNLO{p_{123}}{p_1}{p_{23}} \nonumber \\
&&+\frac{2(d-3)(3d-10)}{(d-4)^2} \frac{1}{(\sab+\sac)\sabc}
\triangleNLO{p_{123}}{p_1}{p_{23}}\nonumber \\
&& -\frac{4(d-3)^2}{(d-4)^2} \frac{1}{(\sab+\sac)\sabc} 
\doublebubblexNLO{p_{123}}{p_{23}}{p_1}\nonumber \\
&& +\frac{4(d-3)^2}{(d-4)^2} \frac{1}{(\sab+\sac)\sabc}\; , 
\doublebubbleNLO{p_{123}}\nonumber \\
&& +\frac{(d-3)(3d-8)(3d-10)}{(d-4)^3} 
\frac{1}{\sbc(\sab+\sac)\sabc} \bubbleNLO{p_{23}} \nonumber \\
&& -\frac{4(d-3)(3d-8)(3d-10)}{(d-4)^3} 
\frac{1}{(\sab+\sac)\sabc^2}
\bubbleNLO{p_{123}}\; ,\\
\triaplanxNLO{p_{123}}{p_{23}}{p_1} &= &-\frac{3(d-3)(3d-10)}{2(d-4)(d-5)}
\frac{1}{(\sab+\sac)^2} \left(\triangleNLO{p_{123}}{p_{23}}{p_1}
+\!\!\trianglexNLO{p_{123}}{p_{23}}{p_1}\right) \nonumber \\
&& +\frac{3(d-3)(3d-8)(3d-10)}{2(d-4)^2(d-5)}\frac{1}{\sbc(\sab+\sac)^2}
\bubbleNLO{p_{23}} \nonumber \\
&& +\frac{3(d-3)(3d-8)(3d-10)}{2(d-4)^2(d-5)}\frac{1}{\sabc(\sab+\sac)^2}
\bubbleNLO{p_{123}} \; .
\end{eqnarray}

For $t=6$, both 
one-loop bubble insertions on propagators of the one-loop box can be reduced:
\begin{eqnarray}
\boxbubbleapNLO{p_{123}}{p_1}{p_2}{p_3} & = & -\frac{3(\sab+\sac)}{\sac\sbc}
\boxbubbleaNLO{p_{123}}{p_1}{p_2}{p_3} \nonumber \\
&&  + \frac{3(3d-10)}{2(d-4)} \frac{1}{\sac\sbc}\left( 
\triangleNLO{p_{23}}{p_2}{p_3}- \triangleNLO{p_{123}}{p_2}{p_{13}}
\right)\nonumber \\
&& -\frac{3(3d-8)(3d-10)}{2(d-4)^2} \frac{1}{\sac^2\sbc}
\bubbleNLO{p_{13}}\; ,\\
\boxbubblebpNLO{p_{123}}{p_1}{p_2}{p_3} & = &
\frac{3}{\sac} \boxbubblebNLO{p_{123}}{p_1}{p_2}{p_3} 
+ \frac{3(3d-10)}{2(d-4)} \frac{1}{\sac\sbc} 
\trianglexNLO{p_{123}}{p_1}{p_{23}} \nonumber \\
&& +\frac{3(3d-8)(3d-10)}{2(d-4)^2} \frac{1}{\sac\sbc\sabc}
\bubbleNLO{p_{123}}\;.
\end{eqnarray}

Among the remaining four planar diagrams at $t=6$, three are reducible to 
simpler subtopologies:
\begin{eqnarray}
\boxxamNLO{p_{123}}{p_1}{p_2}{p_3} & = & -3 \frac{\sab}{\sac\sbc}
\boxxaNLO{p_{123}}{p_1}{p_2}{p_3}  \nonumber \\
&& -\frac{6(d-3)}{d-4}\frac{1}{\sac} 
\boxbubblebNLO{p_{123}}{p_1}{p_2}{p_3} \nonumber \\ 
&& -\frac{3(d-3)(3d-10)}{(d-4)^2} \frac{1}{\sac\sbc}
\trianglexNLO{p_{123}}{p_1}{p_{23}} \nonumber \\ 
&& -\frac{3(d-3)(3d-8)(3d-10)}{(d-4)^3} \nonumber \\
&& \hspace{1cm} \left(\frac{1}{\sac^2\sbc}
\bubbleNLO{p_{13}}+ \frac{1}{\sac\sbc^2} \bubbleNLO{p_{23}}\right)\; , \\
\boxxapNLO{p_{123}}{p_1}{p_2}{p_3} & = & -3 \frac{\sab}{\sac\sbc}
\boxxaNLO{p_{123}}{p_1}{p_2}{p_3}  \nonumber \\
&& +\frac{6(d-3)}{d-4} \frac{\sab+\sac}{\sac\sbc}
\boxbubbleaNLO{p_{123}}{p_1}{p_2}{p_3} \nonumber \\ 
&& -\frac{3(d-3)(3d-10)}{(d-4)^2} \frac{1}{\sac\sbc}
\triangleNLO{p_{23}}{p_2}{p_3} \nonumber \\ 
&& +\frac{3(d-3)(3d-10)}{(d-4)^2} \frac{1}{\sac\sbc}
\triangleNLO{p_{123}}{p_2}{p_{13}} \nonumber \\ 
&& -\frac{3(d-3)(3d-8)(3d-10)}{(d-4)^3} \nonumber \\
&& \hspace{1cm} \left(\frac{1}{\sac\sbc^2}
\bubbleNLO{p_{23}} - \frac{1}{\sac\sbc\sabc} \bubbleNLO{p_{123}}\right)\; , \\
\boxxbmNLO{p_{123}}{p_1}{p_2}{p_3} & = & -\frac{1}{d-4} 
                    \boxxbdotNLO{p_{123}}{p_1}{p_2}{p_3}\nonumber\\
&& + \frac{6(d-3)}{d-4} \frac{\sab+\sac}{\sac\sbc} 
\boxbubbleaNLO{p_{123}}{p_1}{p_2}{p_3} \nonumber \\
&& - \frac{3(d-3)(3d-10)}{(d-4)^2} \frac{1}{\sac\sbc}
\triangleNLO{p_{23}}{p_2}{p_3} \nonumber \\
&& +\frac{3(d-3)(3d-10)}{(d-4)^2} \frac{1}{\sac\sbc}
\triangleNLO{p_{123}}{p_2}{p_{13}} \nonumber \\
&& +\frac{3(d-3)(3d-8)(3d-10)}{(d-4)^3} \frac{1}{\sac^2\sbc} 
\bubbleNLO{p_{13}}\; .
\end{eqnarray}

One of the two remaining non-planar diagrams is also reducible, the other 
non-planar topology contains two master integrals. The reducible 
integral reads:
\begin{eqnarray}
 \boxxbmcrossNLO{p_{123}}{p_1}{p_2}{p_3} & = & 
\hspace{0.28cm}
\frac{3(d-4)}{2d-9}\, \frac{\sab}{\sac\sbc}\boxxaNLO{p_{123}}{p_1}{p_2}{p_3}
\nonumber \\
&&+\frac{3(d-4)}{2d-9}\, \frac{\sbc}{\sab\sac}\boxxaNLO{p_{123}}{p_3}{p_2}{p_1}
\nonumber \\
&&+\frac{3(d-4)}{2d-9}\, \frac{\sac}{\sab\sbc}\boxxaNLO{p_{123}}{p_1}{p_3}{p_2}
\nonumber \\
&&+ \frac{3(d-3)(3d-8)(3d-10)}{(d-4)^2(2d-9)}\, 
\frac{\sac+\sbc}{\sab^2\sac\sbc}\bubbleNLO{p_{12}} \nonumber \\
&&+ \frac{3(d-3)(3d-8)(3d-10)}{(d-4)^2(2d-9)}\, 
\frac{\sab+\sbc}{\sab\sac^2\sbc}\bubbleNLO{p_{13}} \nonumber \\
&&+ \frac{3(d-3)(3d-8)(3d-10)}{(d-4)^2(2d-9)}\, 
\frac{\sab+\sac}{\sab\sac\sbc^2}\bubbleNLO{p_{23}} \nonumber \\
&&- \frac{3(d-3)(3d-8)(3d-10)}{(d-4)^2(2d-9)}\, 
\frac{1}{\sab\sac\sbc}\bubbleNLO{p_{123}}\; .
\end{eqnarray}

\subsection{$t=7$}
At $t=7$ different one finds six different topologies. Three of them 
are triangle insertions to the one-loop box. These three integrals
are all reducible, two of them contain only master integrals up to $t=5$ in 
their reduction:
\begin{eqnarray}
\lefteqn{\boxtriaaNLO{p_{123}}{p_1}{p_2}{p_3}  =}\nonumber \\ 
&& -\frac{6(d-3)(3d-14)}{(d-4)(d-6)}\, \frac{1}{\sbc^2}\,
\boxbubblebNLO{p_{123}}{p_2}{p_1}{p_3}\nonumber \\
&& + \frac{3(3d-14)}{(d-6)} \, \frac{\sab^2}{\sac^2\sbc^2}\,
\boxxaNLO{p_{123}}{p_1}{p_2}{p_3} \nonumber \\
&& -\frac{6(d-3)(3d-14)}{(d-4)(d-6)}\, \frac{(\sab+\sac)^2}{\sac^2\sbc^2}\,
\boxbubbleaNLO{p_{123}}{p_1}{p_2}{p_3} \nonumber \\
&& - \frac{3(d-3)(3d-10)(3d-14)}{2(d-4)^2(d-5)(d-6)} \, \frac{(2d-10)\sab +
  (3d-14) \sbc}{\sac\sbc^2(\sab+\sbc)} \trianglexNLO{p_{123}}{p_2}{p_{13}} 
\nonumber \\
&& + \frac{3(d-3)(3d-10)(3d-14)}{2(d-4)^2(d-5)(d-6)} \, \frac{(2d-10)\sab +
  (3d-14) \sac}{\sac^2\sbc^2} \triangleNLO{p_{23}}{p_2}{p_3} \nonumber\\
&& - \frac{3(d-3)(3d-10)(3d-14)}{2(d-4)^2(d-5)(d-6)} \, \frac{(2d-10)
  \sab\sabc +(3d-14) \sac\sbc }{\sac^2\sbc^2(\sab+\sbc)}
  \triangleNLO{p_{123}}{p_2}{p_{13}} \nonumber \\
&& - \frac{3(d-3)(3d-8)(3d-10)(3d-14)}{2(d-4)^3(d-5)(d-6)} \,
     \frac{(2d-10)\sab + (d-6)\sbc}{\sac^2\sbc^2(\sab+\sbc)} 
   \bubbleNLO{p_{13}}\nonumber\\
&& + \frac{3(d-3)(3d-8)(3d-10)(3d-14)}{(d-4)^3(d-5)(d-6)}\,
\frac{(d-5)\sab - (d-4)\sac}{ \sac^2\sbc^3} \, \bubbleNLO{p_{23}}\nonumber\\
&& + \frac{3(d-3)(3d-8)(3d-10)(3d-14)}{2(d-4)^3(d-5)(d-6)}\, 
     \frac{(2d-10)\sab\sabc + (d-6)
       \sac\sbc}{\sac^2\sbc^2\sabc(\sab+\sbc)}
     \bubbleNLO{p_{123}}\; ,\\
\lefteqn{\boxtriabNLO{p_{123}}{p_1}{p_2}{p_3}  =}\nonumber \\
&& \frac{2(2d-9)}{(d-4)(d-6)}\,\frac{2\sab+\sac+\sbc}{\sac\sbc}
\boxxbdotNLO{p_{123}}{p_1}{p_2}{p_3}\nonumber \\
&& -\frac{3(d-4)}{d-6}\, \frac{(\sac+\sbc)^2}{\sac^2\sbc^2} 
\boxxbNLO{p_{123}}{p_1}{p_2}{p_3}\nonumber \\
&& -\frac{6(d-3)(3d-14)}{(d-4)(d-6)}\, \frac{(\sab+\sbc)^2}{\sac^2\sbc^2} 
\boxbubbleaNLO{p_{123}}{p_2}{p_1}{p_3} \nonumber\\
&& -\frac{6(d-3)(3d-14)}{(d-4)(d-6)}\, \frac{(\sab+\sac)^2}{\sac^2\sbc^2} 
\boxbubbleaNLO{p_{123}}{p_1}{p_2}{p_3} \nonumber\\
&& +\frac{3(d-3)(3d-10)(3d-14)}{2(d-4)^2(d-5)(d-6)}\,
\frac{(2d-10)\sab+(3d-14)\sbc}{\sac^2\sbc^2} \triangleNLO{p_{13}}{p_1}{p_3}
\nonumber \\
&& +\frac{3(d-3)(3d-10)(3d-14)}{2(d-4)^2(d-5)(d-6)}\,
\frac{(2d-10)\sab+(3d-14)\sac}{\sac^2\sbc^2} \triangleNLO{p_{23}}{p_2}{p_3}
\nonumber \\
&& -\frac{3(d-3)(3d-10)}{(d-4)^2(d-6)}\, \frac{(3d-14)\sab + (4d-18)\sac
-(d-4)\sbc }{\sac^2\sbc^2} \triangleNLO{p_{123}}{p_{13}}{p_2}
\nonumber \\
&& -\frac{3(d-3)(3d-10)}{(d-4)^2(d-6)}\, \frac{(3d-14)\sab -(d-4)\sac
+ (4d-18)\sbc }{\sac^2\sbc^2} \triangleNLO{p_{123}}{p_{23}}{p_1}
\nonumber \\
&& -\frac{3(d-3)(3d-8)(3d-10)}{(d-4)^3(d-5)(d-6)} 
\frac{(d-5)(3d-14)(\sab+\sac)+(d-4)^2\sbc}{\sac^3\sbc^2} 
\bubbleNLO{p_{13}}\nonumber \\
&& -\frac{3(d-3)(3d-8)(3d-10)}{(d-4)^3(d-5)(d-6)} 
\frac{(d-5)(3d-14)(\sab+\sbc)+(d-4)^2\sac}{\sac^2\sbc^3} 
\bubbleNLO{p_{23}}\;. 
\end{eqnarray}
The remaining three topologies are the double box and two different 
momentum arrangements of the crossed box. These topologies contain 
each two master integrals. 

\section{Conclusions and Outlook}
\label{sec:conc}
\setcounter{equation}{0}

Progress in the computation of exclusive observables, such as for 
example jet production rates, beyond the next-to-leading 
order has up to now been hampered mainly by difficulties in 
the calculation of virtual two-loop integrals with more than two external 
legs. In contrast to this, many inclusive observables (which correspond 
from the calculational point of view to two-point functions) are
known to next-to-next-to-leading order and even beyond. These 
higher order calculations relied on a variety of elaborate technical tools 
for the computation of the virtual integrals. In this paper, we outline
how techniques known from multi-loop calculations of two-point integrals
 can be modified and extended towards the computation of integrals with a 
larger number of external legs. 

We demonstrate that
the large number of different two-loop integrals appearing in an actual 
calculation can be reduced to a small number of scalar
master integrals by using 
the well-known integration-by-parts identities~\cite{hv,chet} together with 
identities following from Lorentz-invariance which are unique to 
multi-leg integrals. As a by-product of this reduction, one is also able to 
reduce two-loop integrals with tensorial structure to scalar integrals. 
In contrast to two-point integrals, where only a few
topologically different
graphs can appear with potentially large powers of  propagators and 
scalar products, 
one finds that the reduction of three- and four-point integrals gives 
rise to a large number of topologically different graphs, which appear 
however only with small powers of propagators
and scalar products. The reduction of two-point functions usually 
proceeds via solving manually 
the integration-by-parts identities for arbitrary 
powers of propagators and denominators in a given graph topology;
this procedure seems to be not practicable for multi-leg integrals.  
To 
accomplish the reduction 
of these, we developed an algebraic program which automatically
derives and solves the integration-by-parts and Lorentz-invariance 
identities for a given graph up to some pre-selected fixed number of powers in
denominators and scalar products independent of the topology. 

To compute the scalar master integrals, we derive differential equations 
in the external momenta~\cite{remiddi}
for them; the boundary conditions of these 
differential equations correspond to simpler integrals, where for example 
one of the external momenta vanishes.  These differential equations can be 
solved (for arbitrary space-time dimensions) 
by employing standard mathematical methods. We observe that 
the differential equations for the master integrals we considered up to now 
are solved by generalised hypergeometric functions. We illustrate the
application of this method in detail on the example of 
the one-loop four-point function with one off-shell leg. 

Using the differential equation method,
we provide a complete list of all master integrals with up to $t=5$ 
different denominators that can appear in the reduction of two-loop 
four-point functions with one off-shell leg. We also list 
all reducible integrals with $t=6$ and $t=7$ different propagators.
The computation of the master integrals with $t=6$ and $t=7$ is still an 
outstanding task. 

The differential equations for these outstanding master integrals are of 
similar structure as the differential equations for master integrals 
with a smaller number of different propagators. The main problem towards  
a complete computation of these integrals is at present the integration 
of the inhomogeneous term, containing itself already hypergeometric 
functions arising from the subtopologies. 

It is worthwhile to point out similarities and differences between the 
differential equation method employed in this paper and other  
methods employed for similar calculations in the 
literature. Both the negative dimension approach of~\cite{glover} and 
the Mellin-Barnes transformation method employed in~\cite{smirnov,tausk}
rely on choosing a particular assignment of momentum vectors to the 
internal loop propagators. After this assignment, a representation of 
the propagators in terms of a multiple sum (negative dimension approach) 
or an integral transformation (Mellin-Barnes method) is employed, 
such that the integral over the loop momentum can be carried out explicitly. 
The final result for the integral is then retrieved by resummation of 
a multiple sum or by an inverse integral transformation. Both methods, 
when employed for arbitrary space-time dimension, give rise 
to generalised hypergeometric functions, which can be represented as
multiple sums as well as inverse Mellin-Barnes integrals~\cite{bateman}. 
In the differential equation method, one assigns momentum vectors to 
the loop propagators only for the sake of deriving the 
differential equations and the IBP and LI identities.
After applying these identities to simplify the differential equations, 
one obtains a relation between the derivative 
(with respect to an external momentum) of a
master integral, the master integral itself and other master integrals 
with simpler topology, independent of the parametrisation chosen for the 
internal propagators. These differential equations can then be solved 
analytically by
integration; the resulting integrals correspond to the integral 
representations of generalised hypergeometric functions~\cite{bateman}. 
Using the differential equation method, one can therefore circumvent 
the explicit loop momentum integration needed in the other methods and 
one arrives at a representation of the hypergeometric functions, which is 
presumably more transparent than a multiple sum or an inverse integral 
transformation. In the integral representation, it is in particular 
straightforward to identify linear combinations of different 
hypergeometric functions, which are difficult to disentangle in the 
other representations. At present, it should however not be claimed that any 
of the methods is superior, since none of them could yet be 
employed to compute all outstanding two-loop four-point master integrals. 

As a final point, we note that the methods derived in this paper 
contain a 
high level of redundancy, which allows for a number of non-trivial 
checks on the results obtained with them. The automatic reduction to 
master integrals using integration-by-parts and Lorentz-invariance 
identities corresponds to the solution of a linear system of equations 
containing more identities than unknowns. The existence of a solution to 
this system provides therefore already a check on the self-consistency of 
the identities. In computing 
the master integrals from differential equations, one integrates one 
of the three differential equations in the external invariants, such that 
the result can be checked by inserting it in the remaining two differential 
equations. 

In short, this paper demonstrates how techniques developed for 
multi-loop calculation of two-point functions can be extended towards 
integrals with a larger number of external legs. As a first example of 
the application of these tools in practice, we computed some 
up to now unknown 
two-loop four-point functions, relevant for jet calculus beyond the 
next-to-leading order. The most important potential application of these 
tools is the yet outstanding derivation of two-loop virtual corrections to 
exclusive quantities, such as jet observables.

\section*{Acknowledgements}
We are grateful to Jos Vermaseren for his assistance in the use of the 
algebraic program FORM. 
One of the authors (E.R.) wants to thank the Alexander-von-Humboldt
Stiftung for supporting his stay at the Institut f\"ur Theoretische 
Teilchenphysik of the University of Karlsruhe. 
The research work presented in this paper was supported in part by 
the DFG (Forschergruppe ``Quantenfeldtheorie, Computeralgebra und 
Monte-Carlo Simulation'', contract KU 502/8-2).

\begin{appendix}
\renewcommand{\theequation}{\mbox{\Alph{section}.\arabic{equation}}}

\section{Special Functions}
\setcounter{equation}{0}
This appendix summarises the series and integral representations 
of the hypergeometric functions appearing in the master integrals. 
The properties of these functions, in particular their 
region of analyticity, their analytic continuation as well as reduction 
formulae, can be found in the 
literature~\cite{glover,bateman,grad,exton}.

Hypergeometric functions 
are sums with coefficients formed from Pochhammer symbols
\begin{equation}
\left( a\right)_n \equiv \frac{\Gamma(a+n)}{\Gamma(a)}.
\end{equation}

The hypergeometric functions of a single variable are given by:
\begin{eqnarray}
\;_2F_1 \left(a,b;c;z\right) & = & \sum_{n=0}^{\infty} 
\frac{(a)_n(b)_n}{(c)_n} \frac{z^n}{n!}\; ,\\
\;_3F_2 \left(a,b_1,b_2;c_1,c_2;z\right) & = & \sum_{n=0}^{\infty} 
\frac{(a)_n(b_1)_n(b_2)_n}
{(c_1)_n(c_2)_n} \frac{z^n}{n!}\;.
\end{eqnarray}

Two types of  hypergeometric functions of two variables also appear in 
our results:
\begin{eqnarray}
F_1\left(a,b_1,b_2;c;z_1,z_2\right) &=& \sum_{m,n=0}^{\infty} 
\frac{(a)_{m+n}(b_1)_m(b_2)_n}
{(c)_{m+n}} \frac{z_1^m}{m!} \frac{z_2^n}{n!}\; ,\\
S_1\left(a_1,a_2,b;c,d;z_1,z_2\right) &=& \sum_{m,n=0}^{\infty} 
\frac{(a_1)_{m+n}(a_2)_{m+n}(b)_m}
{(c)_{m+n}(d)_m} \frac{z_1^m}{m!} \frac{z_2^n}{n!}\; .
\end{eqnarray}

These functions have the following integral representations:
\begin{eqnarray}
\;_2F_1 \left(a,b;c;z\right) & = & \frac{\Gamma(c)}{\Gamma(b)\Gamma(c-b)} 
\int_0^1 \d t\; t^{b-1} (1-t)^{c-b-1} (1-tz)^{-a} \nonumber \\
&& \hspace{2cm} \mbox{Re}(b) > 0, \qquad \mbox{Re}(c-b)>0 \\
\;_3F_2 \left(a,b_1,b_2;c_1,c_2;z\right) & = & 
\frac{\Gamma(c_1)\Gamma(c_2)}{\Gamma(b_1)\Gamma(c_1-b_1)\Gamma(b_2)
\Gamma(c_2-b_2)}\nonumber \\
&& \int_0^1 \d t_1 \int_0^1 \d t_2\; 
t_1^{b_1-1} t_2^{b_2-1} (1-t_1)^{c_1-b_1-1}(1-t_2)^{c_2-b_2-1}
(1-t_1t_2z)^{-a}\nonumber \\
&& \hspace{-0.7cm}
 \mbox{Re}(b_1) > 0, \qquad \mbox{Re}(c_1-b_1)>0, \qquad 
                \mbox{Re}(b_2) > 0, \qquad \mbox{Re}(c_2-b_2)>0 \\
F_1\left(a,b_1,b_2;c;z_1,z_2\right)&=&\frac{\Gamma(c)}{\Gamma(a)\Gamma(c-a)}
\int_0^1 \d t\; t^{a-1}(1-t)^{c-a-1} (1-tz_1)^{-b_1}(1-tz_2)^{-b_2}
\nonumber \\
&& \hspace{2cm} \mbox{Re}(a) > 0, \qquad \mbox{Re}(c-a)>0 \\
S_1\left(a_1,a_2,b;c,d;z_1,z_2\right) &=& 
\frac{\Gamma(c)\Gamma(d)}{\Gamma(a_1)\Gamma(c-a_1)\Gamma(b)\Gamma(d-b)}
\nonumber \\
&& \int_0^1\d t_1\int_0^1\d t_2 t_1^{a_1-1} t_2^{b-1} (1-t_1)^{c-a_1-1}
(1-t_2)^{d-b-1} \left(1-t_1t_2z_1-t_1z_2\right)^{-a_2} \nonumber\\
&&\mbox{Re}(a_1)>0, \quad  \mbox{Re}(c-a_1)>0,\quad
\mbox{Re}(b)>0, \quad  \mbox{Re}(d-b)>0
\end{eqnarray}

\section{Expansion of Hypergeometric Functions}
\setcounter{equation}{0}

To separate divergent and finite parts 
of the loop integrals derived in this paper, one has to expand them 
around the physical number of space-time dimensions in the parameter 
$\epsilon=(4-d)/2$. We demonstrate in this appendix, that this expansion
can, at least for the hypergeometric functions in one variable, 
be carried out 
in a mechanical way, giving rise to harmonic polylogarithms (HPL),
a generalisation of Nielsen's polylogarithms~\cite{nielsen}
 introduced in~\cite{hpl}. 

Expanding the integral representation of $\,_2F_1$ in $\epsilon$ 
yields simple powers of $(t,1-t,1-tz)$ times the product of some 
number of $(\ln t, \ln (1-t), \ln(1-tz))$. The powers of 
$(t,1-t,1-tz)$ can be integrated by parts until one obtains 
non-trivial integrals 
\begin{displaymath}
\int_0^1 \d t \left(\frac{1}{t},\frac{1}{1-t},\frac{1}{1-tz}\right)
\ln^{n_1}t \ln^{n_2}(1-t) \ln^{n_3}(1-tz)\;. 
\end{displaymath}
All these integrals are combinations of harmonic polylogarithms 
$H(\vec{a};z)$, where $\vec{a}$ is a vector of indices with 
$w=n_1+n_2+n_3+1$ components. $w$ is called the weight of the 
harmonic polylogartihm. 
The proof by induction in $w$ is 
trivial, once the HPL formalism~\cite{hpl} is recalled: 

\begin{enumerate}
\item
Definition of the three HPLs at $w=1$:
\begin{eqnarray}
H(1;z) & \equiv & -\ln (1-z)\; ,\nonumber \\
H(0;z) & \equiv & \ln z\; ,\nonumber \\
H(-1;z) & \equiv & \ln (1+z) 
\label{eq:levelone}
\end{eqnarray}
and the three fractions
\begin{eqnarray}
f(1;z) & \equiv & \frac{1}{1-z} \;, \nonumber \\
f(0;z) & \equiv & \frac{1}{z} \;, \nonumber \\
f(-1;z) & \equiv & \frac{1}{1+z} \; ,
\end{eqnarray}
such that 
\begin{equation}
\frac{\partial}{\partial z} H(a;z) = f(a;z)\qquad \mbox{with}\quad
a=+1,0,-1\;.
\end{equation}
\item For $w>1$:
\begin{eqnarray}
H(0,\ldots,0;z) & \equiv & \frac{1}{w!} \ln^w z\; ,\\
H(a,\vec{b};z) & \equiv & \int_0^z \d x f(a;x) H(\vec{b};x)\; , 
\end{eqnarray}
which results in 
\begin{equation}
\frac{\partial}{\partial z} H(a,\vec{b};z) = f(a;z) H(\vec{b};z)\;.
\end{equation}
This last relation is a convenient tool for verifying identities among 
different HPLs. Such identities can be verified by first checking a 
special point (typically $z=0$)
and subsequently checking the derivatives. If agreement in 
the derivatives is not obvious, this procedure can be repeated until one 
arrives at relations involving only HPLs with $w=1$.
\item
The HPLs fulfil an algebra (see Section 3 of~\cite{hpl}), such that 
a product of two HPLs (with weights $w_1$ and $w_2$) 
of the same argument $z$ is a combination of HPLs of argument 
$z$ with weight $w=w_1+w_2$. 
\end{enumerate}

Using these properties of the HPL, one can show that 
the integrals appearing in the $\epsilon$-expansion of the 
hypergeometric function can be reexpressed as 
\begin{equation}
\int_0^1 \d t \left(\frac{1}{t},\frac{1}{1-t},\frac{1}{1-tz}\right)
\ln^{n_1}t \ln^{n_2}(1-t) \ln^{n_3}(1-tz) \to 
\int_0^1 \d t \left(\frac{1}{t},\frac{1}{1-t},\frac{1}{t-1/z}\right)
H(\vec{a},t) H( \vec{b},zt)\;.
\label{eq:NTint}
\end{equation}
Following the argumentation of Section 7 of~\cite{hpl}, one can 
show that the integral on the right hand side of the above equation 
yields a linear combination of 
 HPLs of weight $w=w_a+w_b+1$. The proof goes via induction in $w_b$.

For $w_a=w_b=0$ one has $ H(\vec{a};t) H( \vec{b};zt)=1$. 
The $t$-integral in (\ref{eq:NTint}) yields then a combination 
of HPL of weight $w=1$
(\ref{eq:levelone}). Likewise, for $w_b=0$ the right hand side of 
(\ref{eq:NTint}) will yield a linear combination of HPLs
of weight $w=w_a+1$ and of argument $z$, as proven in Section 7 of~\cite{hpl}. 

Considering 
\begin{equation}
\frac{\partial}{\partial z}
\int_0^1 \d t \left(\frac{1}{t},\frac{1}{1-t},\frac{1}{t-1/z}\right)
H(\vec{a};t) H(B, \vec{b};zt)\;,
\label{eq:induct}
\end{equation} 
we observe that
\begin{eqnarray}
\frac{\partial}{\partial z} \frac{\d t}{t-1/z} &=& \frac{1}{z^2}
\, \frac{1}{t-1/z}\, \d t\,
 \frac{\partial}{\partial t} + \mbox{boundary terms} \;,
\nonumber \\
\frac{\partial}{\partial z} H(B,\vec{b};zt) &=& t\,f(B;zt)\,H(\vec{b};zt)
\nonumber \;.
\end{eqnarray}
Making these replacements in (\ref{eq:induct}) and applying 
partial fractioning to all denominators, we are left with
\begin{equation}
\left(\frac{1}{z},\frac{1}{1-z},\frac{1}{1+z}\right)
\int_0^1 \d t \left(\frac{1}{t},\frac{1}{1-t},\frac{1}{t-1/z}\right)
H(\vec{a};t) H(\vec{b};zt)\;, 
\end{equation}
which is a combination of HPLs with argument $z$ and
weight $w=w_a+w_b+1$ multiplied with 
$(1/z,1/(1-z),1/(1+z))$. Integrating 
(\ref{eq:induct}) over $z$ will thus yield a combination of HPLs with 
argument $z$ and weight $w+1$, which completes the proof by induction.

The $\epsilon$-expansion of $\,_3F_2$, corresponding 
to a double integral in $t_1$ and $t_2$, is obtained by carrying out the 
procedure described here twice, again resulting in a combination 
of HPLs. 
A systematic $\epsilon$-expansion of $F_1$ and $S_1$, which are functions  
of two variables $z_1$ and $z_2$, will in general go beyond the harmonic 
polylogarithms in one variable.

\end{appendix}

\end{document}